\documentstyle[aps,epsf]{revtex}
\def\PRL#1#2#3{Phys. Rev. Lett. {\bf #1}, #2 (#3)}
\def\be{\begin{equation}}
\def\ee{\end{equation}}
\def\bea{\begin{eqnarray}}
\def\eea{\end{eqnarray}}
\def\nn{\nonumber\\}
\def\r#1{(\ref{#1})}
\def\up{\uparrow}

\makeatletter
\def\preprint#1{\def\@preprint{
\noindent\hfill\hbox{#1}\vskip10pt}}
\makeatother
\begin{document}
\draft
\preprint{OUTP-98-12S}
\title{Faceting Transition in an Exactly Solvable
Terrace-Ledge-Kink Model}
\vspace{1.5 em}
\author{Douglas B. Abraham\cite{present},
Fabian H. L. E\char'31ler and
Franck T. Latr\'emoli\`ere}
\address{Department of Physics, Theoretical Physics, Oxford
University,\\
1 Keble Road, Oxford, OX1 3NP, United Kingdom}
\date{\today}
\maketitle
\begin{abstract}
We solve exactly a Terrace-Ledge-Kink (TLK) model describing a
crystal surface at a microscopic level. We show that there is a
faceting transition driven either by temperature or by the
chemical potential that controls the slope of the surface.
In the rough phase
we investigate thermal fluctuations of the surface using
Conformal Field Theory.
\end{abstract}

\section{Introduction}

In this paper, we reexamine a model of crystal surfaces in three
dimensions \cite{dba}, which is inspired by the
Terrace-Ledge-Kink (TLK) \cite{burton} ideas of Kossel and Stransky
\cite{kossel}. By careful specification, an isomorphism can be
established between our system and an extension of the six-vertex
models \cite{6v}, which allows an exact discussion at a thermodynamic
level of the different types of phase transitions which take place. We
complement this with recent relevant deductions from conformal
field theory,
which give microscopic insight. We shall return to this six-vertex
isomorphism after the next few paragraphs, in which we describe the
model and outline the results.

Consider a vicinal section of a crystal surface, that is one
which on
average is tilted by a small angle from a closed packed plane.
Let
this underlying plane have normal $(001)$ and let the normal to
the mean
vicinal surface lie in the plane containing $(001)$ and $(101)$.
Following Kossel and Stransky
\cite{kossel}, we give the vicinal surface a
microscopic
structure by regarding it as the upper surface of a set of unit
cubes,
each of which representing a molecule or atom, which are stacked
vertically with no voids, thereby covering the basal plane
$(x,y,0)$
with columns of cubes.
We shall make a somewhat eccentric
choice of lattice structure, which is described in detail
below, and which will turn
out to be of crucial importance in the
statistical-mechanical modelling of the
short-ranged interactions between ledges.

We now impose the vicinal condition by making the height on the
left extremum of the
surface equal to zero, and the height on the right one equal to
some constant
positive integer. These extrema intersect the lines $(0,y)$ and
$(M,y)$
of $\Lambda$ as zig-zags or zippers, as shown in
Fig.~\ref{fig:fig1}.
Connected components of the upper surface having the same height
are
termed terraces and are separated from other terraces by ledges,
which
can have bends through an angle of $\pi/2$: these are termed
kinks.

In this paper, we
shall make two further restrictions: firstly, there
are no
adatoms or pits on the terraces; and secondly, ledges cannot
separate
adjacent terraces with a height difference greater than unity.
This
means that different ledges can never coincide and that no ledge
can
form a closed loop: each ledge begins at the bottom of the
lattice and
exits at the top.

We emphasize that the above construction
is not the only way of generating exactly solvable models.
Other routes are the BCSOS \cite{bcsos,eberlein}
and the surface
deconstruction \cite{deconstruction} model.
There is also the Hamiltonian limit type of approximation
proposed by Jayaprakash et al. \cite{jaya}
and by Burkhardt and Schlottmann \cite{burk},
together with
other work by Villain et al. \cite{villain}
and by Bartelt et al. \cite{bartelt}.
There is a general review of morphology of vicinal
surfaces in \cite{science}.  For small tilt angles
$\theta$,
a polynomial approximation is given to the surface
free energy $f(\theta)$ where
\be
f(\theta)=f(0)+a\theta+b\theta^3
\label{approx}
\ee
in which $a$ is related to the kink energy
and $b$ comes from ledge-ledge interactions
which can be entropic in character,
elastic (although the distinction between these two is
not entirely clear) or coming from electrostatic
dipolar interactions in metals.
We scrutinize \r{approx} from our exact
solution and also analyse the ledge-ledge interaction
from first principles in appendix E.

Our model shows the formation of a crystal
facet at half-filling, which we can regard as at
a tilt angle of $\pi/4$.  In the low-temperature
phase, the surface height fluctuations about the
$\pi/4$ tilted plane are of order one, but above the
faceting temperature there is power-law decay of
ledge-ledge correlation functions, making
Kosterlitz-Thouless \cite{kt} behaviour extremely
likely.  As we shall see, for a certain
{\sl nonzero}
repulsion between ledges we get a free-fermionic structure
of the transfer matrix,
and in this case the height fluctuations are shown
to be of Kosterlitz-Thouless type.

Finally, we note that our model is closely related to a lattice
regularization of the anisotropic principal chiral field \cite{PCF}.

\section{The model}

\begin{figure}[ht]
\begin{center}
\noindent
\epsfxsize=0.4\textwidth
\epsfbox{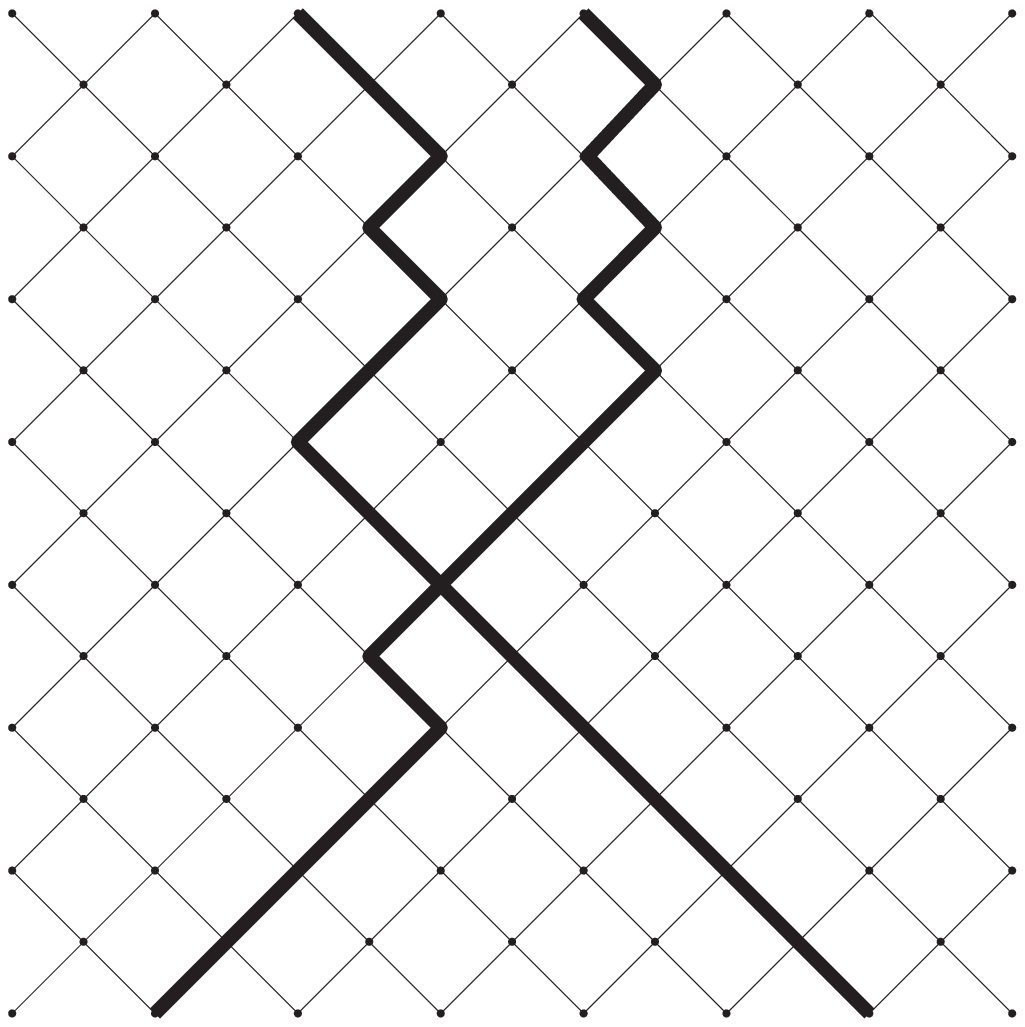}
\hfill
\epsfxsize=0.4\textwidth
\epsfbox{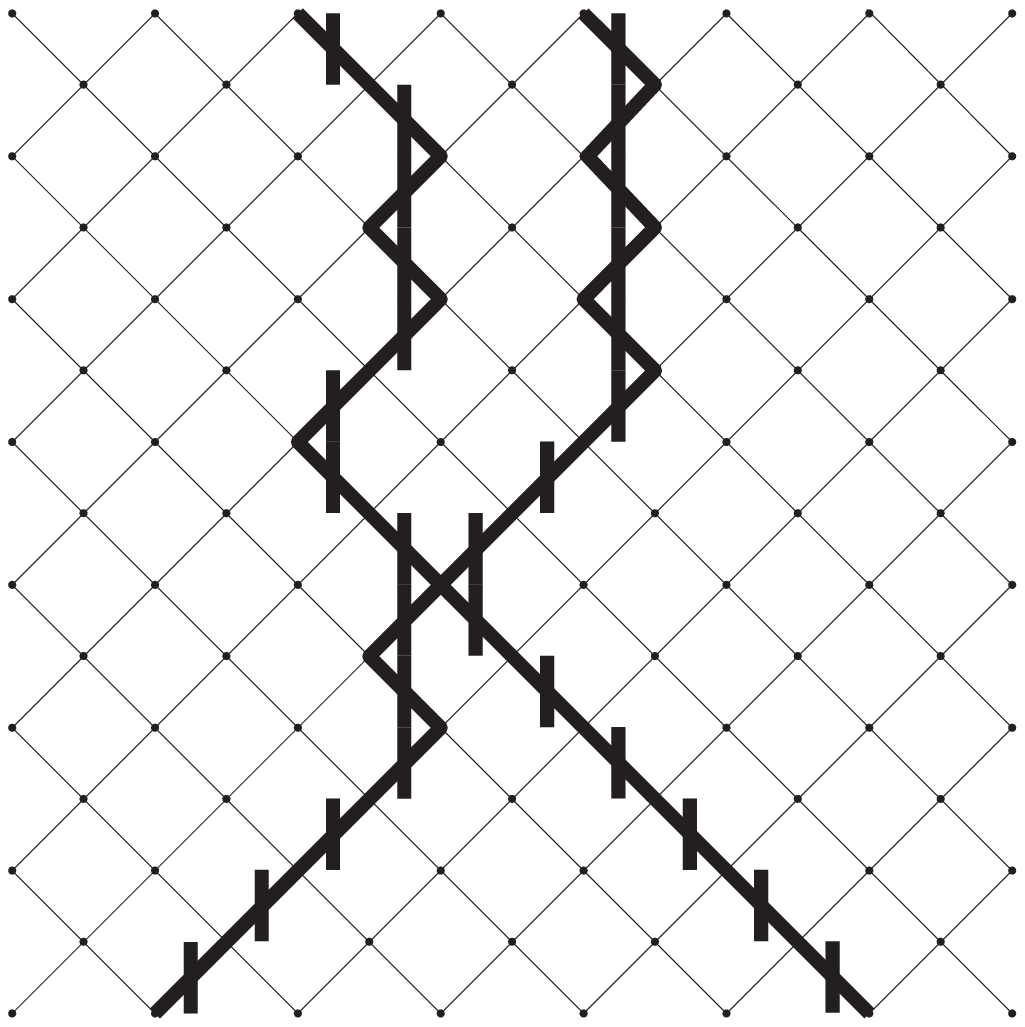}
\end{center}
\caption{\label{fig:fig2}
(a) six-vertex model configuration;
(b) definition of ledge segments
for a given six-vertex configuration.}
\end{figure}

We consider a square lattice tilted by 45 degrees (see
Fig.~\ref{fig:fig2}
(a)), and define a configuration by declaring each link between
two lattice
sites to be occupied (state up, denoted 1) or empty (state down,
denoted 2). We then impose the
six-vertex constraint or ice rule: at each vertex (centered at
lattice
sites) the number of occupied links must be conserved between
top and
bottom. An example of an allowed configuration is shown in
Fig.~\ref
{fig:fig2}(a), where the occupied links are
drawn as thick
lines.

\begin{figure}[ht]
\begin{center}
\noindent
\epsfxsize=0.4\textwidth
\epsfbox{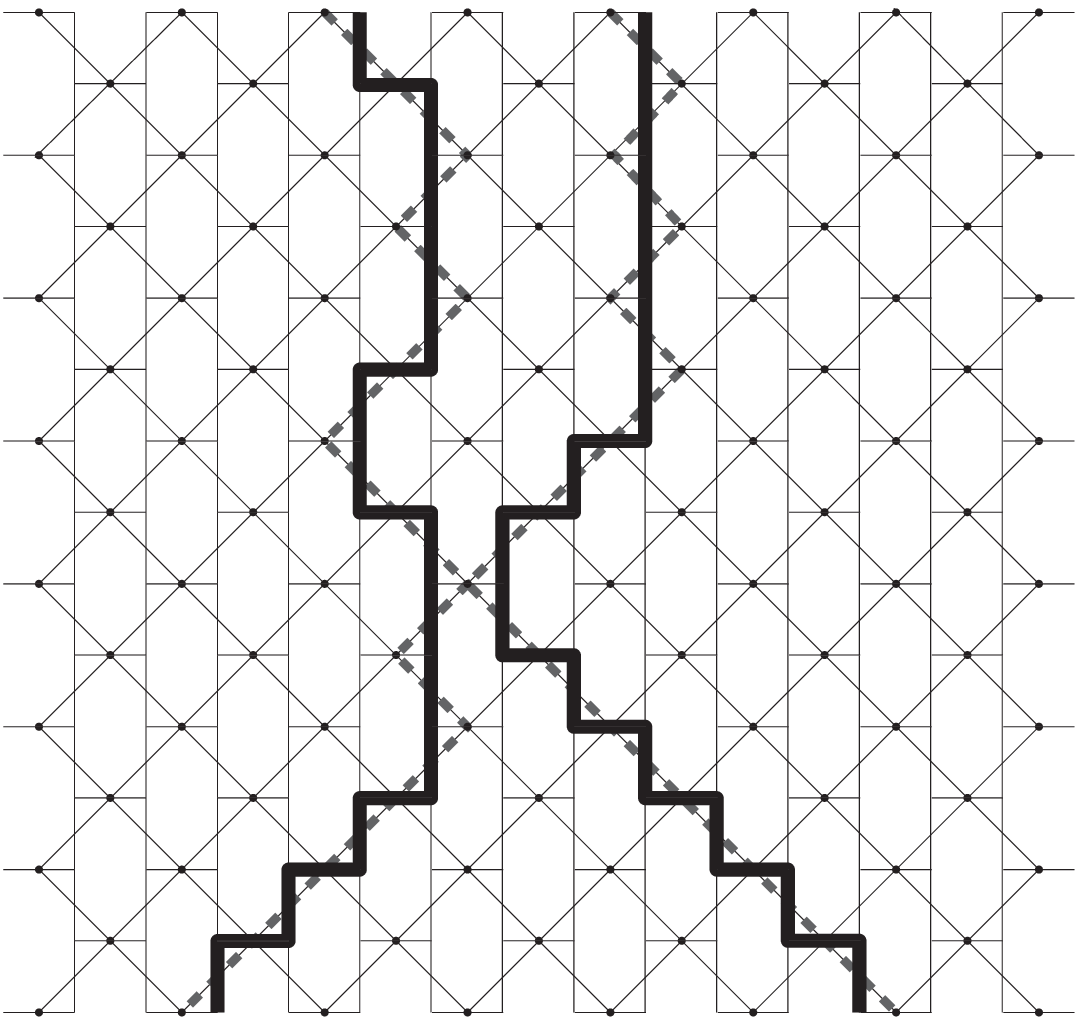}
\hfill
\epsfxsize=0.4\textwidth
\epsfbox{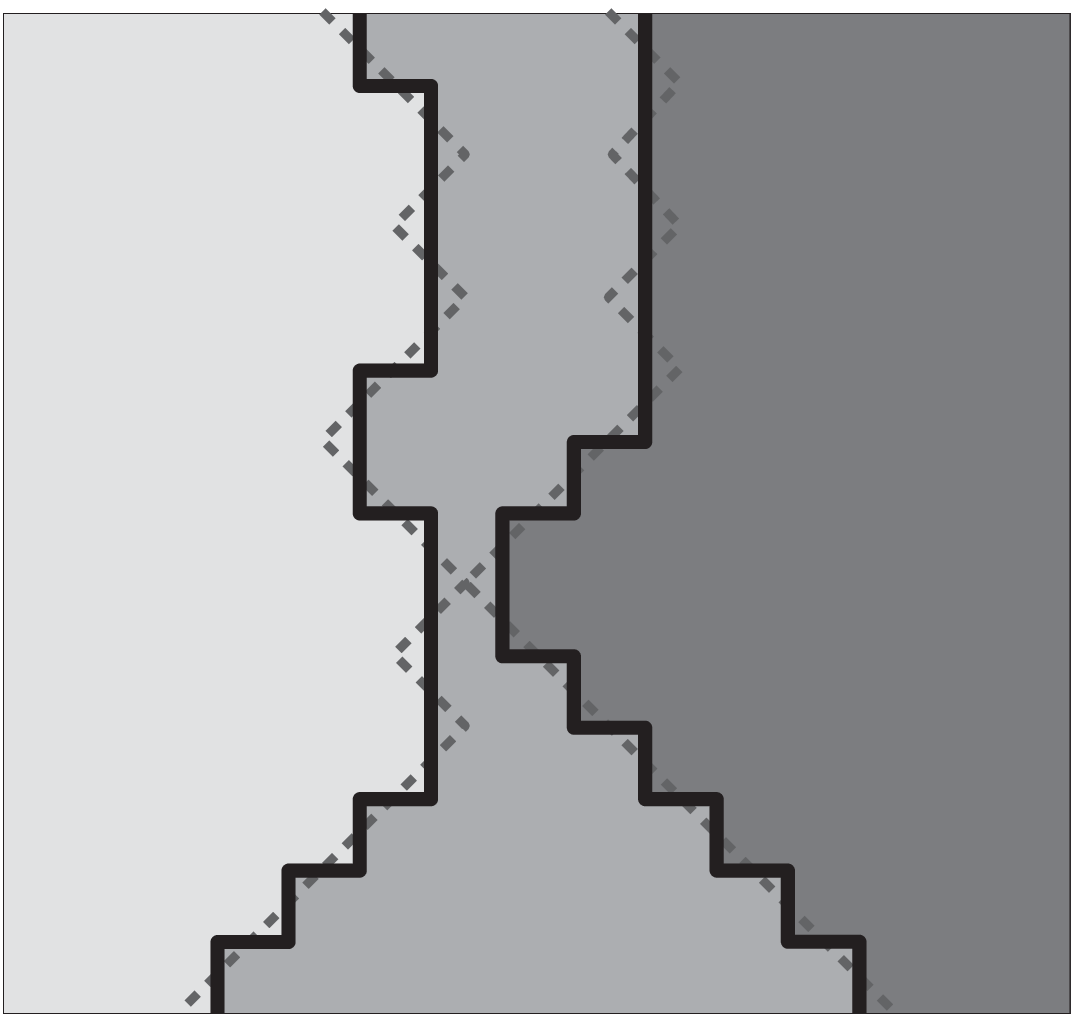}
\end{center}
\caption{\label{fig:fig2c}
Square lattice configurations in the TLK model:
(a) ledges and kinks; (b) terraces.}
\end{figure}

Every such allowed configuration can now be mapped onto a
configuration of
terraces, ledges and kinks defining a surface. We first
introduce {\sl ledges}
by drawing short vertical lines through the middle of each
occupied link
(see Fig.~\ref{fig:fig2}(b)).
These segments are then connected by horizontal lines
representing {\sl kinks},
as shown in Fig.~\ref{fig:fig2c}(a). Treating ledges and kinks
as steps
on a surface, the above procedure then defines {\sl terraces},
which are indicated by different shades of grey in
Fig.~\ref{fig:fig2c}(b).
Figure ~\ref{fig:fig1} explicits the three-dimensional
structure.

The above procedure leads to a number of restrictions on the
allowed
configurations of terraces, ledges and kinks. For example
ledges cannot form closed loops,
and there cannot be
multiple kinks.

\begin{figure}[ht]
\begin{center}
\noindent
\epsfxsize=0.4\textwidth
\epsfbox{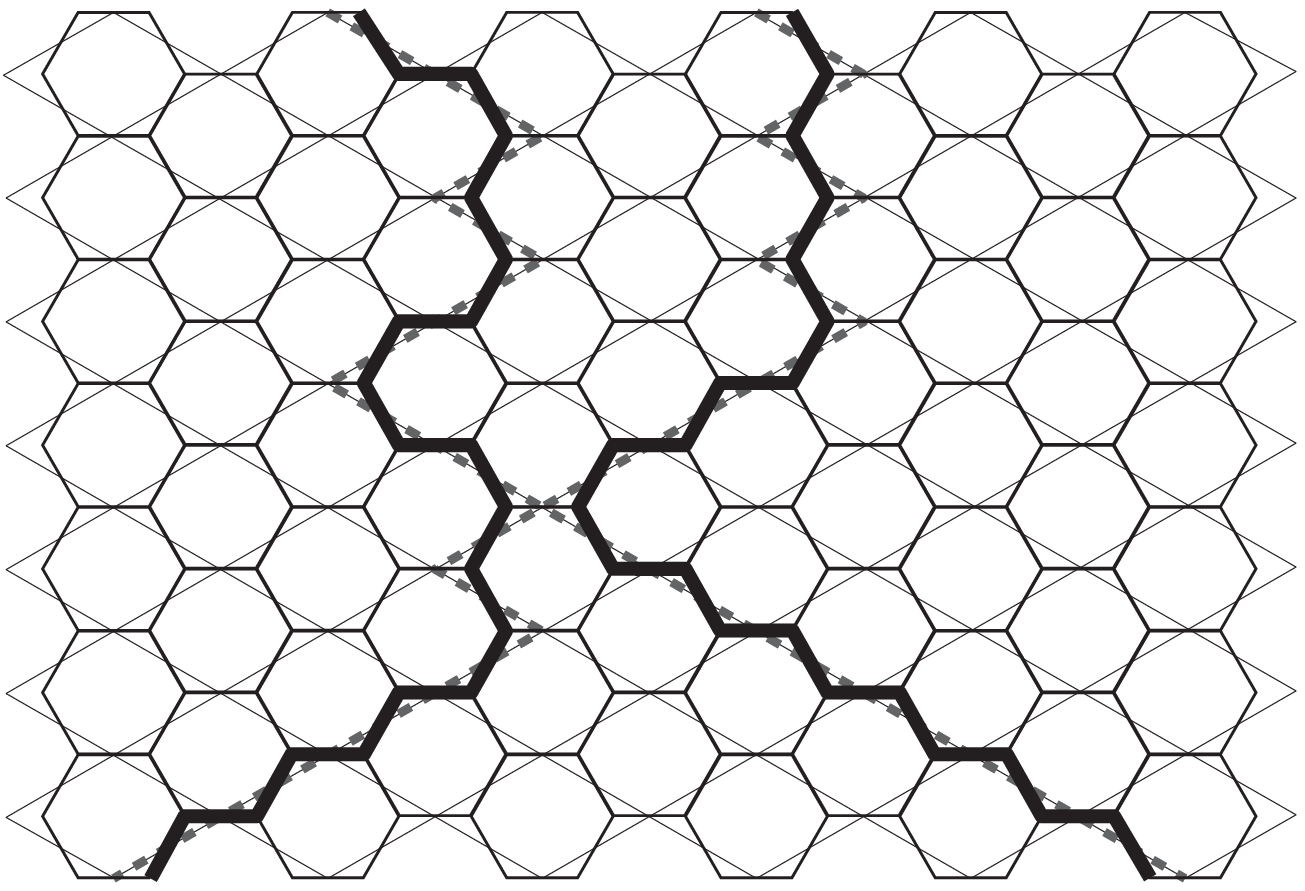}
\hfill
\epsfxsize=0.4\textwidth
\epsfbox{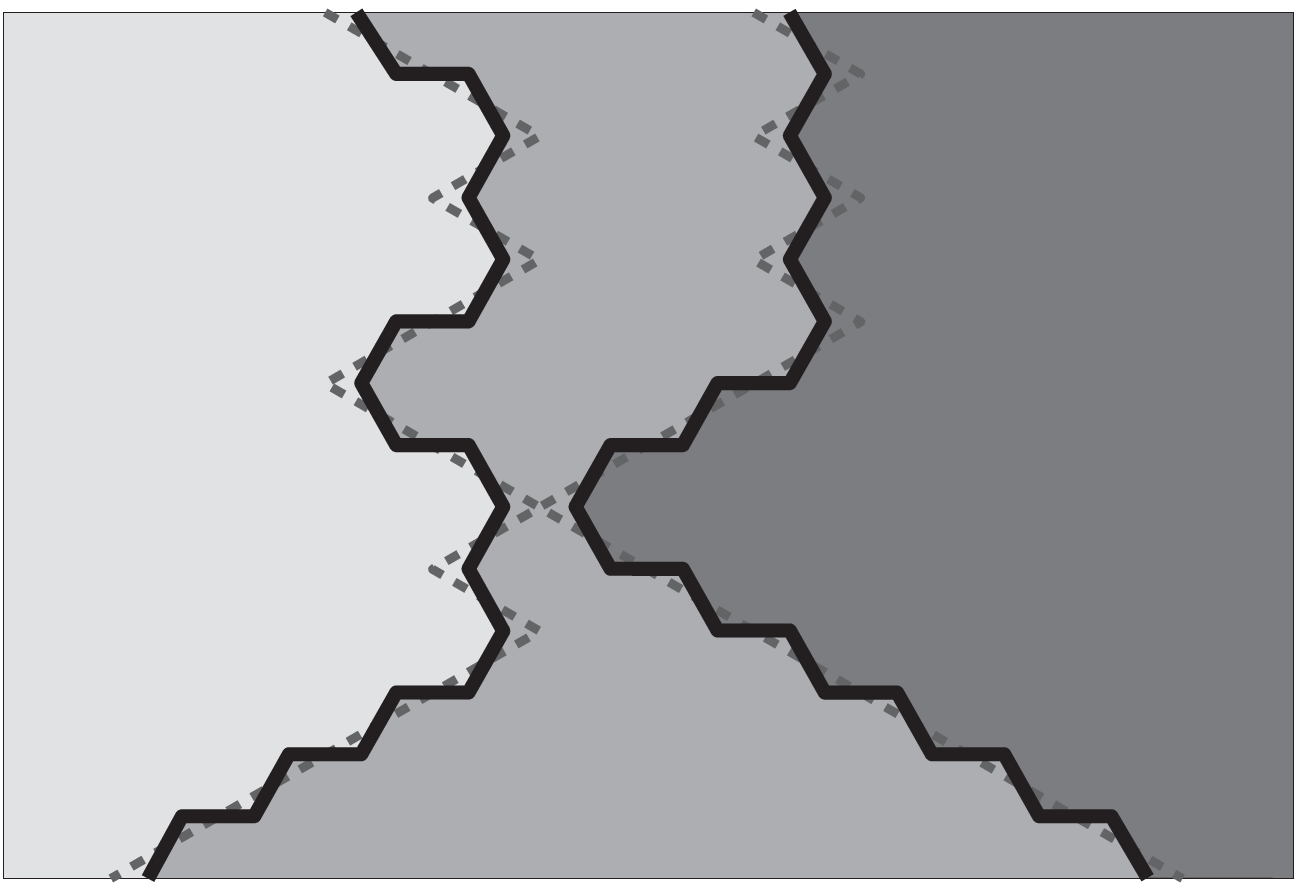}
\end{center}
\caption{\label{fig:fighex2c}
A configuration in the TLK model with the hexagonal lattice.}
\end{figure}

This last restriction can be effectively removed
by a slight modification of the ledge representation.
Stretching the lattice by $\sqrt{3}$ in the
horizontal
direction we can introduce a lattice of regular hexagons, and use
the
sides of these hexagons to define the ledges and kinks as shown
in Fig.~\ref{fig:fighex2c}.
In this case
the restrictions on multiple kinks are built into the lattice
structure.

\begin{figure}[ht]
\begin{center}
\noindent
\epsfxsize=0.45\textwidth
\epsfbox{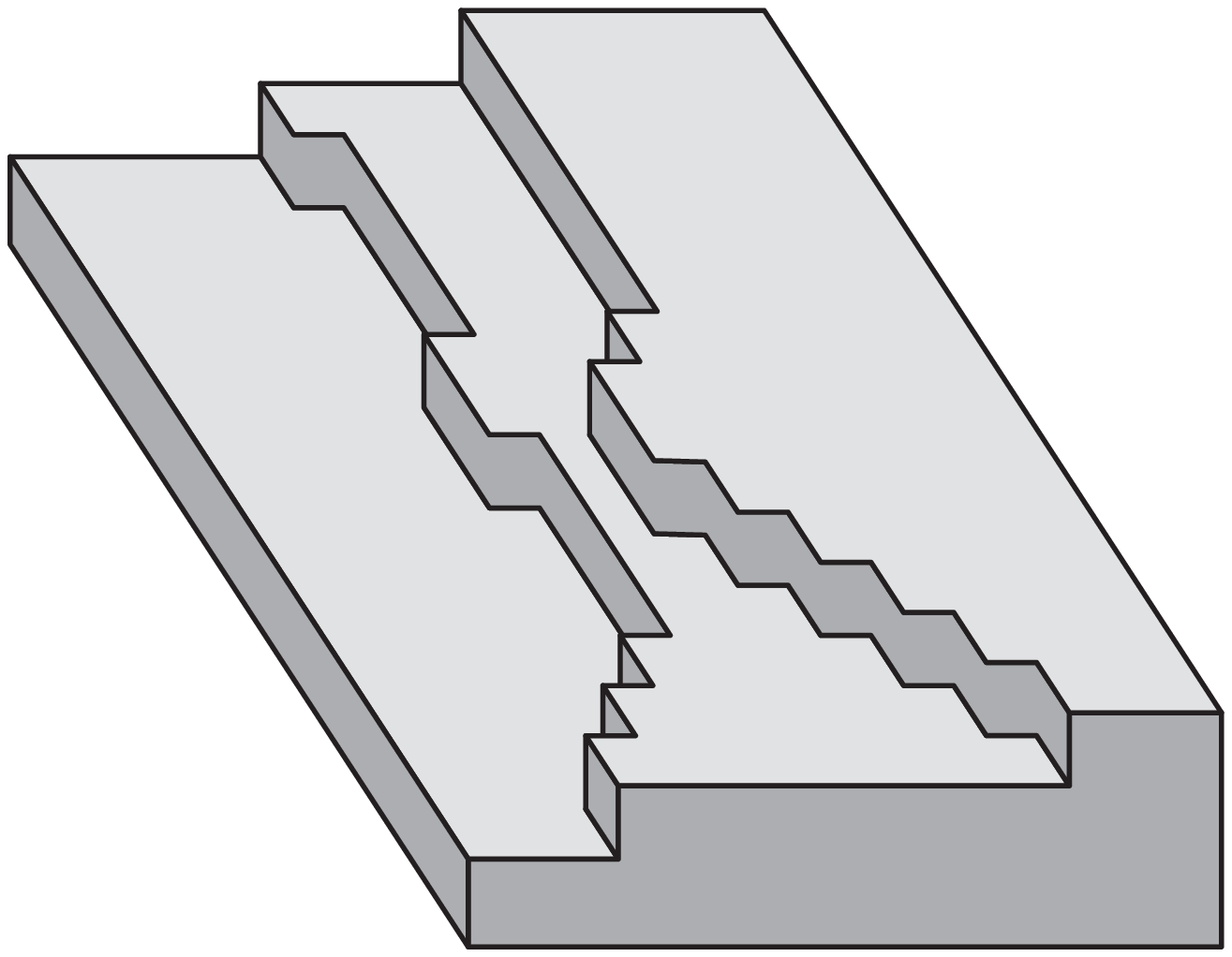}
\hfill
\epsfxsize=0.45\textwidth
\epsfbox{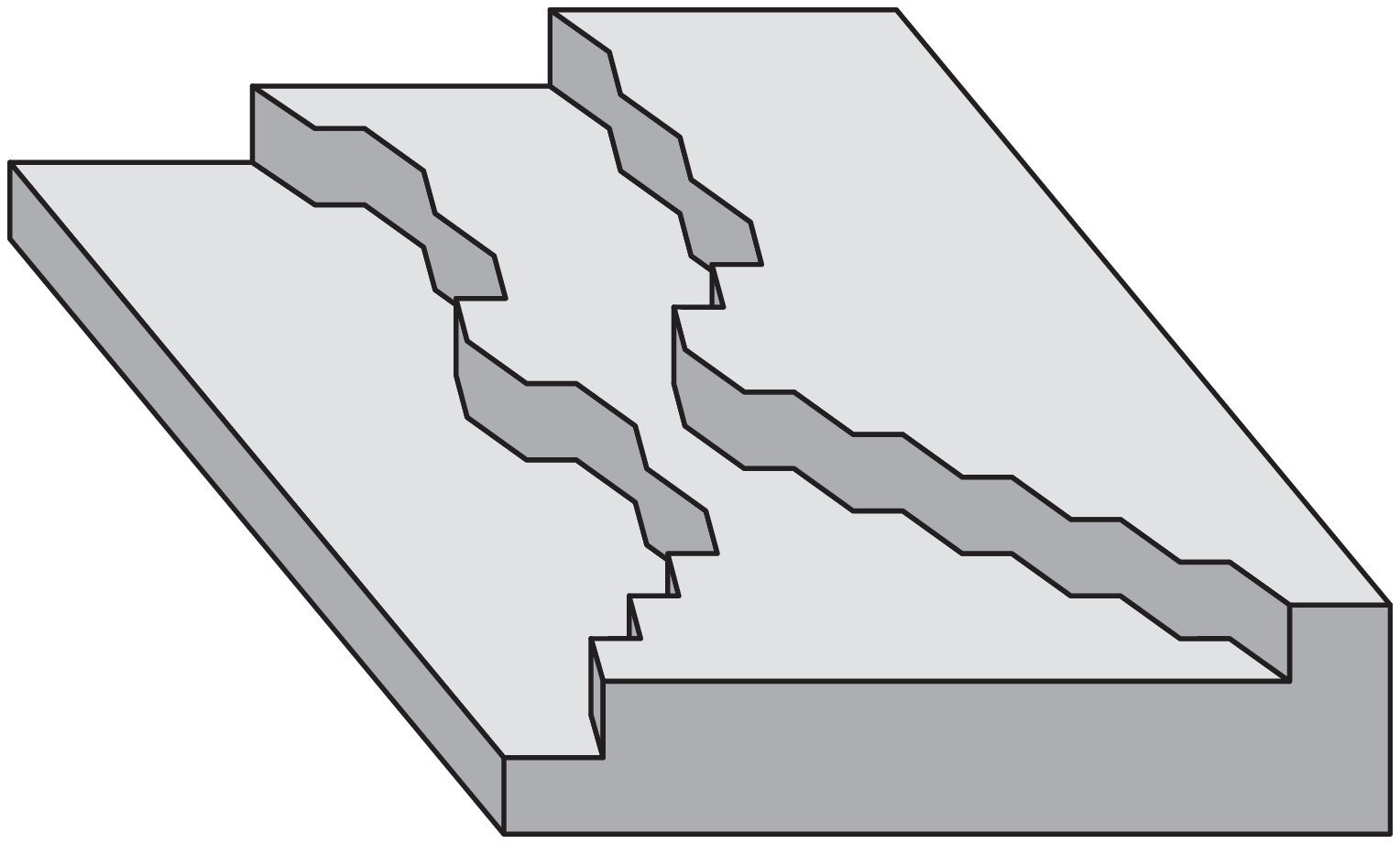}
\end{center}
\caption{\label{fig:fig1}
Section of a crystal surface: (a) square lattice;
(b) hexagonal lattice.}
\end{figure}

We therefore end up with
two slightly different geometrical representations
for the configurations: in Fig.~\ref{fig:fig1} (a) the underlying
lattice is square, although there are restrictions in the
allowed configurations which involve the tiling
of Fig.~\ref{fig:fig2c};
in Fig.~\ref{fig:fig1} (b) the
terrace-ledge-kink structure is placed on a hexagonal lattice.

We now associate a weight with each configuration.

\begin{figure}[ht]
\begin{center}
\noindent
\epsfxsize=0.9\textwidth
\epsfbox{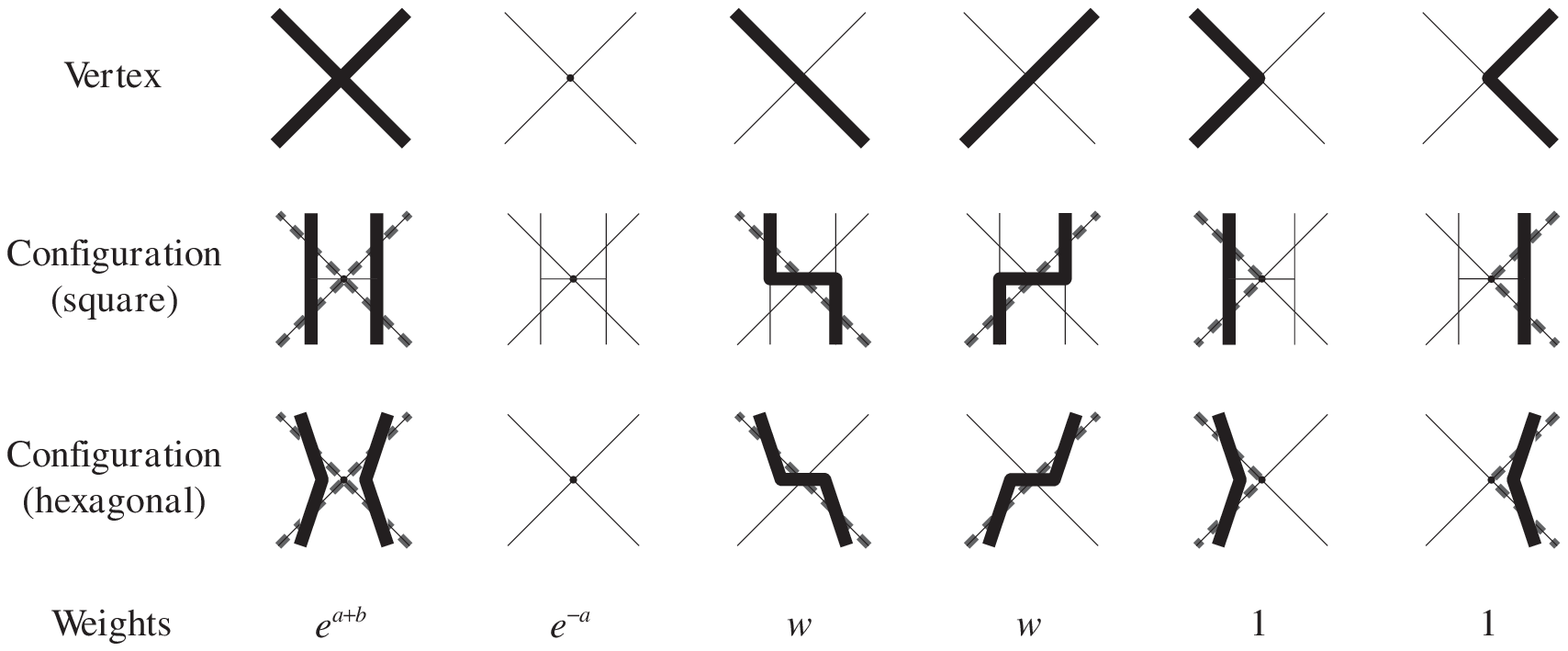}
\end{center}
\caption{\label{fig:figver}
Six-vertex vertices and the two possible interpretations
in terms of ledges.}
\end{figure}

In terms of the
underlying six-vertex model we associate a Boltzmann weight
according to the
rule shown in Fig.~\ref{fig:figver} with each vertex. The weight
$w$ is a kink fugacity,
and lower values of $w$ correspond to stiffer ledges.
The weights also depend on
the interaction energy $b$ between neighbouring ledges and the
ledge stiffness
(note that in our model only neighbouring ledges
interact and that
by construction ledges can never cross each other).
The parameter $a$ is a ledge fugacity or chemical potential,
which controls the number of ledges, or equivalently
the slope of the crystal surface.

\begin{figure}[ht]
\begin{center}
\noindent
\epsfxsize=0.9\textwidth
\epsfbox{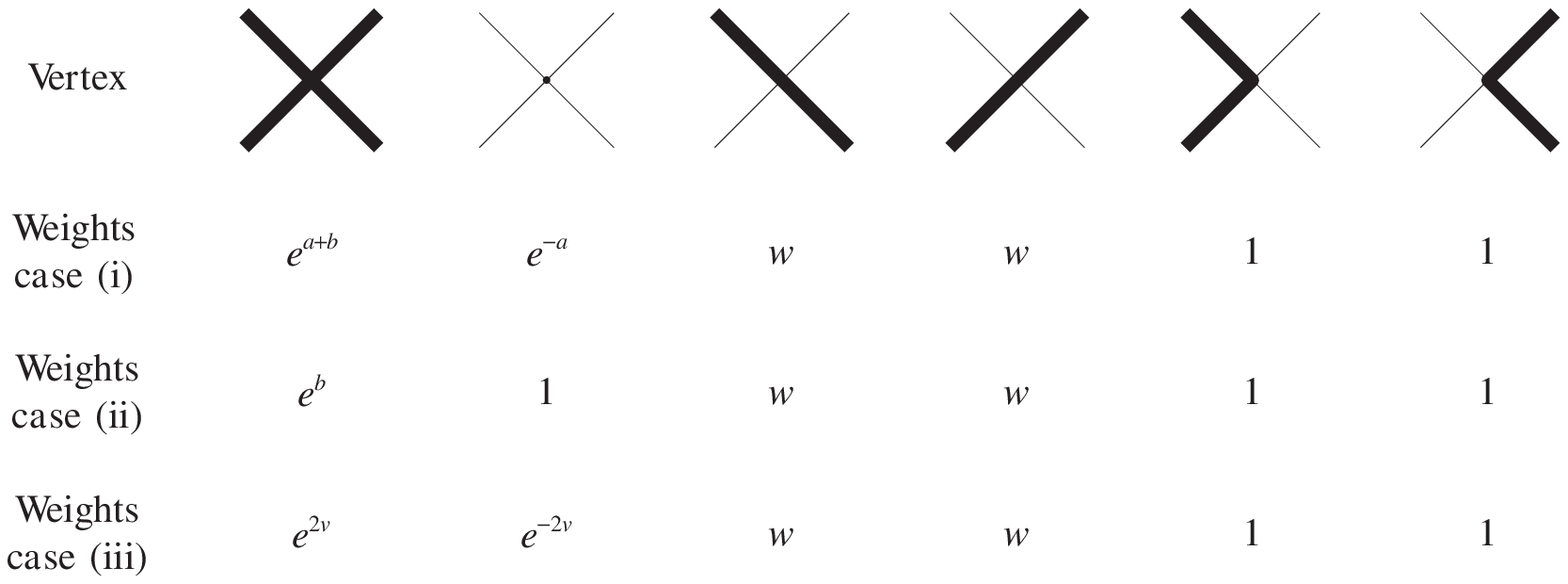}
\end{center}
\caption{\label{fig:figweights}
The three different assignments
of vertex weights discussed in the text.}
\end{figure}

We now discuss the relationships between the three weight
assignments shown on Fig.~\ref{fig:figweights}.
Some aspects of the Bethe Ansatz for
(ii) have been discussed in
\cite{dba}. The Bethe Ansatz generates eigenvectors
for the $(1,1)$ transfer matrix with fixed fraction
of up arrows, corresponding to a given vicinal section
angle. The solution gives the limiting free energy
per unit area, in principle for all angles, but in closed
form by Fourier methods only for the half-filled case.
It was argued that there
are two quite distinct types of phase transition,
which occur at the solutions of
$e^b=(1\pm w)^2$.

For $b>0$ (attraction between ledges), there is a reconstructive
phase transition at a temperature $T(b)$ to a state with zero
limiting entropy for $T<T(b)$. As the temperature approaches
$T(b)$ from above, the specific heat diverges as
$(T-T(b))^{-1/2}$ (cf. end of
appendix E; this is similar to \cite{sph}).

In this paper, we focus on the transition at
$e^b=(1-w)^2$, for $b<0$, implying repulsion between the ledges.
This problem was discussed in the fixed vertical polarization
case with the $(1,1)$ transfer matrix in \cite{dba}.
There it was found that
at half-filling the system displays an infinite-order
phase transition similar to the Kosterlitz-Thouless one
\cite{kt}, or the van Beijeren one in the BCSOS case \cite{bcsos}.

We consider the weight assignment
(i) of Fig.~\ref{fig:figweights}
and we use a Legendre transform with respect to the variable
$a$, with no fixing of the polarization, in order to go to
the physical situation of case (ii)
with fixed polarization. Notice that
neither (i) nor (ii) is equivalent to the usual $(h,v)$
field scenario in the six-vertex model (iii):
new results are needed.
We note that the model (i) is not the same as (ii) because the
chemical potential $a$ is not
always an invertible function of the
ledge number. In other words, the model \cite{dba} contains phases
that are not accessible by tuning $a$ in the model considered
here. The situation is analogous to the six-vertex model in a vertical
field or equivalently the Heisenberg XXZ chain in a magnetic field
$H$. If we consider the ferromagnet in the sector with fixed
magnetization the ground state is rather complicated \cite{aks}. 
This phase is not accessible by tuning the magnetic field and
considering the absolute ground state ({i.e.} the minimal energy
state in the grand canonical ensemble where the number of up/down
spins is variable) as the ground state will always be the saturated
ferromagnetic state with all spins up or down.

We
use the Quantum Inverse Scattering Method,
which is reviewed in \cite{vladb}, to derive new results,
as well as the old ones in a formulation which may
be more familiar for most readers.
We calculate:
\be
E(a,b)=\lim_{\Lambda\to\infty}\frac{1}{|\Lambda|}
\log Z_\Lambda(a; b, w)
\ee
where
\be
Z_\Lambda(a; b, w)={\rm Trace}\prod_{j=1}^{|\Lambda|}
e^{bn_1}e^{a(n_1-n_2)}w^{n_3+n_4}
\ee
where the trace is taken over all ice-rule-compliant
configurations on $|\Lambda|=MN$ points for a
$M\times N$ lattice; $n_j$ is the number of vertices of
type $j=1,2,\dots,6$ in the lattice configuration.
The free energy for fixed polarization is then given by
\be
F(m; b, w)=\sup_a(E(a, b)-ma).
\label{legendre}
\ee
Correlation functions may be calculated in either
ensemble by appropriately relating $m$ and the optimal
$a={\hat a}(m)$ in \r{legendre}, which is unique provided
that $\partial E/\partial a$ exists at $a={\hat a}(m)$.

The paper is structured as follows. In section III, we
give the Bethe Ansatz equations and expressions for
the eigenvalues of the transfer matrix.
A detailed derivation of these results,
as well as the classification of the model
in the framework of the Quantum Inverse
Scattering Method, appears in appendix A.
In section IV we discuss the phase diagram and
calculate explicitly the free energy. 
In section V we give a precise definition of height
and height-difference correlation functions.
Sections VI and VII are concerned with the determination
of the large-distance behaviour of these correlation
functions. Finally we summarize our results and
discuss their relevance in relation
to other theoretical and experimental work.

There are several further appendices discussing
some technical aspects of our work: the free-fermion
case is solved explicitly in appendix B, and
appendix C deals with the analysis of the Bethe Ansatz
equation in the case of negative kink energy,
where novel peculiarities concerning the distribution
of roots in the complex plane are encountered.
Appendix D gives the second-quantized
form of the transfer operator, and in appendix E
we discuss effective interactions between ledges
and some low-angle expansions.

\section{Bethe Ansatz equations}

Denoting the weight matrix by $R$ we have
in case (i):
\begin{equation}
R^{11}_{11}=e^{a+b}\ ,\qquad R^{22}_{22}=e^{-a}\ ,\qquad
R^{12}_{21}=w\ ,\qquad R^{21}_{12}=w\ ,\qquad R^{12}_{12}=1\ ,
\qquad R^{21}_{21}=1\ .
\label{weights}
\end{equation}

The vertex model defined by (\ref{weights}) is exactly solvable
by Bethe
Ansatz. In \cite{dba} a particular parametrization of the Bethe
Ansatz equations (for the case $a=0$) was derived using coordinate
space techniques. In appendix A we show that the model
(\ref{weights}) can actually be embedded into a family
of commuting transfer matrices and derive a parametrization of
the Bethe Ansatz equations in terms of entire functions. In the
language of the Quantum Inverse Scattering Method \cite{vladb} the
model (\ref{weights}) is simply a particular case of an inhomogeneous
asymmetric 6-vertex model \cite{bethe,6v,a6v}. As is shown in appendix
A, it is convenient to introduce parameters $\gamma$, $\omega$ and $v$
such that
\begin{equation}
e^{b}=\frac{\sin^2(\gamma-\omega/2)}{\sin^{2}\gamma},\qquad
w=-\frac{
\sin\omega/2}{\sin\gamma},\qquad
e^{-a}=e^{-2v}\frac{\sin(\gamma-\omega/2)}{
\sin\gamma}. \label{para}
\end{equation}
Eigenstates of the transfer matrix (with $2N-M$ ledges) are
parametrized in
terms of rapidity variables $\lambda_k$, which are subject to
the following
set of coupled algebraic equations, called Bethe Ansatz equations
\begin{equation}
\left( \frac{\sinh\left(\lambda_k+i\frac{\gamma}{2}\right)} {
\sinh\left(\lambda_k-i\frac{\gamma}{2}\right)}\
\frac{\sinh\left(\lambda_k-i
\frac{\omega-\gamma}{2}\right)}
{\sinh\left(\lambda_k-i\frac{\omega+\gamma}{2
}\right)}\right)^N= -\prod_{j=1}^M
\frac{\sinh\left(\lambda_k-\lambda_j+i{
\gamma}\right)}
{\sinh\left(\lambda_k-\lambda_j-i{\gamma}\right)}\ ,\
k=1\ldots M\ . \label{baeq}
\end{equation}
The eigenvalues of the operator of translation by two sites are
given by
\begin{equation}
\exp\left(iP(\{\lambda_j\})\right)=\prod_{k=1}
\frac{\sinh\left(\lambda_k-i
\frac{\gamma}{2}\right)}
{\sinh\left(\lambda_k+i\frac{\gamma}{2}\right)}\
\frac{\sinh\left(\lambda_k-i\frac{\omega+\gamma}{2}\right)} {
\sinh\left(\lambda_k-i\frac{\omega-\gamma}{2}\right)}\ ,
\label{transl}
\end{equation}
and the eigenvalues $\Lambda_{dd}$ of the diagonal-to-diagonal
transfer
matrix $\tau_{dd}$ are
\begin{equation}
\Lambda_{dd}(\{\lambda_j\})=e^{2a(M-N)}
\left(\frac{\sin(\gamma-\frac{\omega
}{2})}{\sin\gamma}\right)^{2M}
\prod_{j=1}^M\frac{\sinh\left(\lambda_j+i
\frac{\gamma}{2}\right)}
{\sinh\left(\lambda_j-i\frac{\gamma}{2}\right)}
\frac{\sinh\left(\lambda_j-i\frac{\gamma+\omega}{2}\right)} {
\sinh\left(\lambda_j+i\frac{\gamma-\omega}{2}\right)}\ .
\end{equation}

\section{Phase diagram}

The parametrization (\ref{para}) is such that $\gamma$ and
$\omega$ cannot
always chosen to be real. Furthermore, due to the fact that $a$
and $b$ must
be real and $w$ must be real and between
$0$ and $1$ there are restrictions on
the regions
of ``physical'' $\gamma$ and $\omega$. As $a$ and $b$ are real
so is $v$,
and by exchanging up and down spins in the Bethe Ansatz solution
(see appendix A) we can restrict ourselves to the case $v<0$.

The permitted values of $b$ and $w$ are covered by the following three
regimes of $\gamma$ and $\omega$:

\begin{itemize}
\item {\underbar{Region~(1a):} } $\gamma $ and $\omega $ are real
with
$0<\gamma<\pi/2$ and $-2\gamma<\omega<0$;

\item {\underbar{Region~(1b):} } $\gamma $ and $\omega $ are real
with
$\pi/2<\gamma<\pi$ and $-2\pi+2\gamma<\omega<0$;

\item {\underbar{Region~(2):} } $\gamma $ and $\omega $ are purely
imaginary. It is
convenient to choose a parametrization obtained from
(\ref{para}) and (\ref{baeq})
by taking $\gamma \to i\gamma $, $\omega \to i\omega $
and
$\lambda \to
i(\lambda +\pi /2)$, which leads to the following set of
equations

\begin{equation}
e^{b}=\frac{\sinh ^{2}(\gamma -\omega /2)}{\sinh ^{2}\gamma
},\qquad w=-
\frac{\sinh \omega /2}{\sinh \gamma },\qquad
e^{-a}=e^{-2v}\frac{\sinh
(\gamma -\omega /2)}{\sinh \gamma },
\end{equation}
where $\gamma >0$ and $-2\gamma<\omega <0$,
\begin{equation}
\left( \frac{\cos \left( \lambda _{k}+i\frac{\gamma }{2}\right)
}{\cos
\left( \lambda _{k}-i\frac{\gamma }{2}\right) }\ \frac{\cos
\left( \lambda
_{k}-i\frac{\omega -\gamma }{2}\right) }{\cos \left( \lambda
_{k}-i\frac{
\omega +\gamma }{2}\right) }\right)
^{N}=-\prod_{j=1}^{M}\frac{\sin \left(
\lambda _{k}-\lambda _{j}+i{\gamma }\right) }{\sin \left( \lambda
_{k}-\lambda _{j}-i{\gamma }\right) }\ ,\ k=1\ldots M\ .
\label{bae2}
\end{equation}
\begin{equation}
\Lambda _{dd}(\{\lambda _{j}\})=e^{2a(M-N)}\left( \frac{\sinh
(\gamma -\frac{
\omega }{2})}{\sinh \gamma }\right)
^{2M}\prod_{j=1}^{M}\frac{\cos \left(
\lambda _{j}+i\frac{\gamma }{2}\right) }{\cos \left( \lambda
_{j}-i\frac{
\gamma }{2}\right) }\frac{\cos \left( \lambda _{j}-i\frac{\gamma
+\omega }{2}
\right) }{\cos \left( \lambda _{j}+i\frac{\gamma -\omega
}{2}\right) }\ .
\label{evr2}
\end{equation}

\item {\underbar{Region~(3):} } $\omega $ and $\gamma -\pi$ are
purely imaginary. A convenient parametrization is then
\begin{equation}
e^{b}=\frac{\sinh ^{2}(\gamma -\omega /2)}{\sinh ^{2}\gamma
},\qquad w=\frac{
\sinh \omega /2}{\sinh \gamma },\qquad e^{-a}=e^{-2v}\frac{\sinh
(\gamma
-\omega /2)}{\sinh \gamma },
\end{equation}
where $2\gamma>\omega>0$.
We substitute $\lambda\to i(\lambda+\pi/2)$.
The Bethe Ansatz equations become
\begin{equation}
\left( \frac{\sin \left( \lambda _{k}+i\frac{\gamma }{2}\right)
}{\sin
\left( \lambda _{k}-i\frac{\gamma }{2}\right) }\ \frac{\sin
\left( \lambda
_{k}-i\frac{\omega -\gamma }{2}\right) }{\sin \left( \lambda
_{k}-i\frac{
\omega +\gamma }{2}\right) }\right)
^{N}=-\prod_{j=1}^{M}\frac{\sin \left(
\lambda _{k}-\lambda _{j}+i{\gamma }\right) }{\sin \left( \lambda
_{k}-\lambda _{j}-i{\gamma }\right) }\ ,\ k=1\ldots M\ .
\label{bae3}
\end{equation}
The eigenvalues of $\tau _{dd}$ are
\begin{equation}
\Lambda _{dd}(\{\lambda _{j}\})=e^{2a(M-N)}\left( \frac{\sinh
(\gamma -\frac{
\omega }{2})}{\sinh \gamma }\right)
^{2M}\prod_{j=1}^{M}\frac{\sin \left(
\lambda _{j}+i\frac{\gamma }{2}\right) }{\sin \left( \lambda
_{j}-i\frac{
\gamma }{2}\right) }\frac{\sin \left( \lambda _{j}-i\frac{\gamma
+\omega }{2}
\right) }{\sin \left( \lambda _{j}+i\frac{\gamma -\omega
}{2}\right) }\ .
\label{evr3}
\end{equation}
\end{itemize}

By analysing the Bethe Ansatz equations in the various regions
we can
determine the largest eigenvalues of the transfer matrix and
thus establish
the phase diagram of the model.

\subsection{Phase diagram for $v=0$}

\begin{figure}[ht]
\begin{center}
\noindent
\epsfxsize=0.4\textwidth
\epsfbox{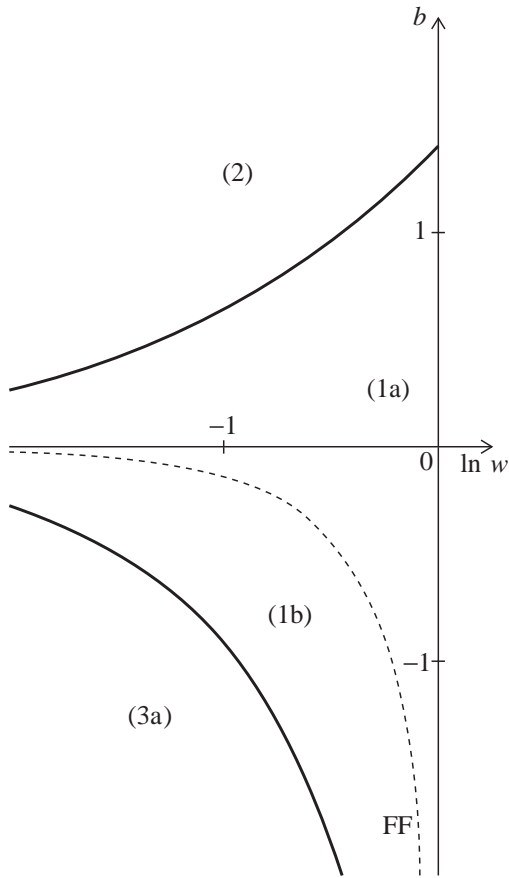}
\end{center}
\caption{\label{fig:8}
Phase diagram for zero vertical field $v=0$.}
\end{figure}

As in the case of the symmetric six-vertex model the phase
structure of
the TLK model for vanishing vertical field $v=0$ is closely
related to the
parametrization of the Bethe Ansatz equations. We find that
there are three
different phases corresponding to the three different regimes in
$\gamma$
and $\omega$ discussed above. Phase~1 is a critical (rough)
phase whereas
Phase~2 and Phase~3 correspond to massive phases (smooth surfaces).
They are the
analogs of the ferro (2) and antiferroelectric (3) phases in the
symmetric
six-vertex model. The resulting phase diagram for $v=0$ is shown
in Fig.~\ref
{fig:8}.

The precise parametrization of the phase boundaries for $v=0$ is
as follows:
the boundary between Region~(1)
and Region~(2) corresponds to $\gamma$ and
$\omega$ small with
$\omega/\gamma$ finite, and
\begin{equation}
e^{b}=(1+w)^{2} \qquad{\rm with}\qquad
w=-\frac{\omega}{2\gamma}\ .
\end{equation}
The boundary between Region~(1) and
Region~(3) is for $\gamma-\pi$ and $\omega$
small with $\omega/(\pi-\gamma)$ finite
\begin{equation}
e^{b}=(1-w)^{2} \qquad{\rm with}\qquad
w=-\frac{\omega}{2(\pi-\gamma)}\ .
\end{equation}

\subsection{Phase diagram for $v<0$}

We now discuss the phase diagram as a function of $v$ for
various points in the $(\gamma,\omega)$ plane. The situation is
analogous to the one for the XXZ Heisenberg chain in a magnetic
field
\cite{yaya}. As far as the parametrization of the Bethe Ansatz
equations is concerned, the $(\gamma,\omega)$ plane is divided
into
Regions~(1)-(3) as discussed above. At $v=0$ the phase structure of
the
model exactly reflects this division of the $(\gamma,\omega)$
plane.

\begin{itemize}
\item {} Let us consider a fixed point $P_{1}=(\gamma _{1},\omega
_{1})$ in
Region~(1). At $v=0$ the model is thus critical. As $v$ is
decreased the phase
boundary between Phase~1 and Phase~2 moves in negative $b$ direction
and
eventually encompasses $P_{1}$: the model is then massive and we
are in the
ferroelectric phase. $P_{1}$ remains in Phase~2 as $v$ is
decreased further.

\item {} Next we consider a point $P_{2}=(\gamma
_{2},\omega_{2})$ in
Region~(2). For any value of $v\leq 0$ the model is in Phase~2.

\item {} Finally we consider a point $P_{3}=(\gamma _{3},\omega
_{3})$ in Region~(3). At $v=0$ the model is in the massive Phase~3.
As $v$ decreases the phase boundary between Phase~1 and Phase~3 is
shifted in
the negative $b$ direction and eventually $P_{3}$ enters Phase~1.
As $v$ is decreased further, $P_{3}$ finally enters into Phase~2.
\end{itemize}

\begin{figure}[ht]
\begin{center}
\noindent
\epsfxsize=0.4\textwidth
\epsfbox{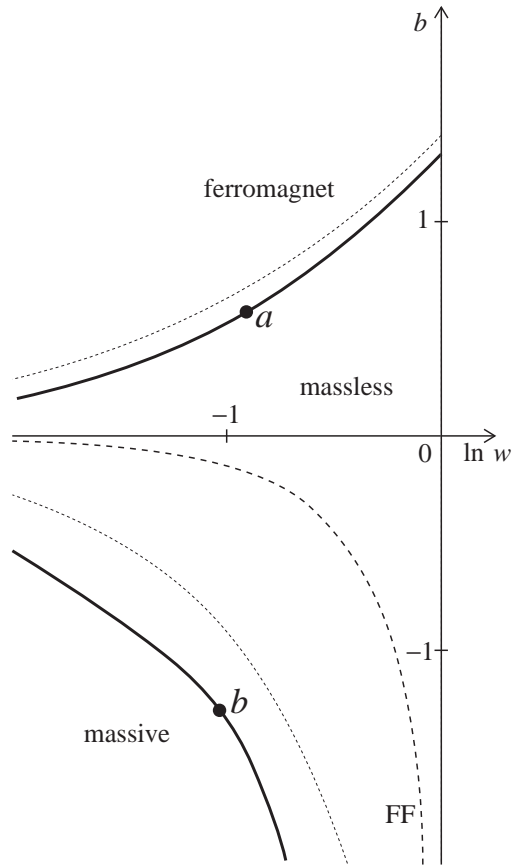}
\end{center}
\caption{\label{fig:9}
Phase diagram for $v=-0.025$ (solid lines) vs $v=0$
(dotted lines).}
\end{figure}

The above picture, together with the precise parametrization of
the
phase boundaries for $v<0$, follows from the analysis of the
largest
eigenvalues of $\tau_{dd}$, which is outlined in the next
subsection. The change of the phase boundaries under a decrease
of $v$
is shown in Fig.~\ref{fig:9}. The corresponding ledge chemical
potential tends to decrease the number of ledges in the system
and as
a result the transition to the state with no ledges (Phase~2)
occurs
at a smaller value of ledge attraction $b$ for fixed $w$. On the
other
hand the critical regime extends to stronger ledge repulsion.

\subsection{Characterization of the Maximal Eigenvalue}

Let us now turn to the distribution of rapidities $\lambda$
corresponding to
the largest eigenvalue of $\tau_{dd}$ in the three phases. It is
convenient
to define an ``energy'' $E$ by
\begin{equation}
E(\{\lambda_k\})=-\ln|\Lambda_{dd}(\{\lambda_k\})|
=2Na+\sum_{k=1}^M\epsilon_0(\lambda_k) \ ,
\end{equation}
where the ``bare energy'' of a single ledge
$\epsilon_0(\lambda_k)$ is given
explicitly below. The maximal eigenvalue of $\tau_{dd}$ then
corresponds to
the minimal energy $E$. In the following we call the state
corresponding to the largest eigenvalue of $\tau_{dd}$ the
``ground state''
of the system.

\begin{itemize}
\item {\underbar{Region~(1):}} Let us start with the
parametrization for the
Bethe Ansatz equations and eigenvalues of the transfer matrix of
Region~(1).
The contribution of a single ledge to the energy and to the
eigenvalue of
the 2-site translation operator (\ref{transl}) are
\begin{eqnarray}
\epsilon _{0}(\lambda ) &=&-\log \left| \frac{\sinh \left(
\lambda +i\frac{
\gamma }{2}\right) \sinh \left( \lambda -i\frac{\gamma +\omega
}{2}\right) }{
\sinh \left( \lambda -i\frac{\gamma }{2}\right) \sinh \left(
\lambda +i\frac{
\gamma -\omega }{2}\right) }\right| -4v\ , \nonumber \\
e^{ip_0(\lambda )} &=&\frac{\sinh (\lambda +i\gamma /2)\sinh
(\lambda -i\omega
/2+i\gamma /2)}{\sinh (\lambda -i\gamma /2)\sinh (\lambda
-i\omega
/2-i\gamma /2)}\ . \label{eps0}
\end{eqnarray}
In the following we restrict ourselves to solutions of the
Bethe Ansatz
equations for which
\begin{equation}
\left| e^{ip_0(\lambda _{k})}\right| =1,\ k=1\ldots M.
\label{nostring}
\end{equation}
This means that we do not consider so-called string solutions
\cite{taka,bethe,strings} as they are not important for our
present
purposes: the largest few eigenvalues of $\tau _{dd}$ are given
by solutions of (\ref{baeq}) that fulfill (\ref{nostring}).

Rapidities fulfilling (\ref{nostring}) are distributed on the
two lines $\lambda =x+i\pi /2+i\omega /4$
and $\lambda =x+i\omega /4$. We
find that $\epsilon _{0}(x+i\pi /2+i\omega /4)<0$ and
$\epsilon _{0}(x+i\omega /4)>0$
for any real $x$. Therefore for $v<0$ the distribution of
rapidities in the
ground state is obtained by filling a Fermi sea of rapidities on
the line $\lambda =x+i\pi /2+i\omega /4$ with $|x|\leq x_{F}$.

The Fermi sea disappears if $\epsilon _{0}(i\omega /4)>0$ and
the maximum
eigenvalue state then contains no ledges at all. Thus for
sufficiently large
negative $v$ we are in the ferroelectric Phase~2. We
exclude this case
from the following discussion.

In the thermodynamic limit the ground state is described by a
continuous
rapidity distribution function $\rho (x)$, which is defined by
\begin{equation}
\rho (x_{k})=\lim_{N\to \infty }1/(N(x_{k+1}-x_{k}))\ ,
\end{equation}
where $\lambda _{k}=x_{k}+i\pi /2+i\omega /4$ are solutions of
(\ref{baeq}).
The Bethe Ansatz equations turn into an integral equation for
$\rho (x)$ in
the thermodynamic limit, which is easily obtained by
subtracting the logarithm of
(\ref{baeq})
for $k$ and $k+1$ and then turning sums into integrals
\begin{equation}
\rho (x)-\frac{1}{2\pi }\int_{-x_{F}}^{x_{F}}dy\ K(x-y)\rho
(y)=\rho _{0}(x),
\label{rhointeq}
\end{equation}
where
\begin{eqnarray}
K(x) &=&-\frac{2\sin 2\gamma }{\cosh 2x-\cos 2\gamma }\ ,
\nonumber \\
\rho _{0}(x) &=& \frac{1}{2\pi}\frac{dp_0}{dx}
=\frac{1}{\pi }\left( \frac{\sin (\gamma +\omega /2)}{\cosh
2x+\cos (\gamma +\omega /2)}+\frac{\sin (\gamma -\omega
/2)}{\cosh 2x+\cos
(\gamma -\omega /2)}\right) .
\label{Krho01}
\end{eqnarray}
The ground state energy density is then given by
\begin{equation}
E_{GS}=\int_{-x_{F}}^{x_{F}}dy\ \rho(y)\
\epsilon _{0}(y+i\pi /2+i\omega/4)\ ,
\label{Egs2}
\end{equation}
where $x_{F}$ is determined self-consistently by the condition
\begin{equation}
\frac{\delta E_{GS}}{\delta x_{F}}\bigg|_{v}=0.
\end{equation}
The density of ledges in the ground state as a function of $v$
is then given by
\begin{equation}
D=\int_{-x_{F}}^{x_{F}}dx\ \rho (x)\ .
\label{ledgedens}
\end{equation}
Note that $D=1$ corresponds to one ledge per {\sl two} sites,
{i.e.}
half-filling because there are $2N$ sites in $x$-direction.
In the square lattice representation,
if we assume that the step height corresponding to a ledge
is equal to two lattice steps,
then the tilt angle of the surface with ledge density
$D$ is given by $\tan\theta=D$.  The
angle corresponding to half filling
is then $\theta=\pi/4$, and we denote
by $\theta_f=\arctan(2)$ the angle for full filling.

It is easier
to work with the so-called dressed energy $\epsilon (x)$
defined by
\be
\epsilon (x)-\frac{1}{2\pi }
\int_{-x_{F}}^{x_{F}}dy\ K(x-y)\ \epsilon (y)
=\epsilon _{0}(x+i\pi /2+i\omega /4) ,
\label{dresseden}
\ee
where we demand that $\epsilon (x_{F})=0$. This fixes $x_{F}$ as
a function
of $v$ and it can be shown that
\begin{equation}
E_{GS}=\int_{-x_{F}}^{x_{F}}dy\ \rho _{0}(y)\ \epsilon (y)\ .
\label{Egs}
\end{equation}
In order to calculate the ground state energy one now simply
solves (\ref{dresseden}) and (\ref{Egs}) numerically (analytic
solutions are not possible for generic $v<0$). As remarked above
we are dealing with a critical phase.
In order to facilitate the
calculation of correlation functions we need to determine the
finite-size correction to the ground state energy, which is
related
\cite{centralc} to the central charge of the underlying conformal
field theory \cite{cft}. The leading corrections are determined
by using the Euler-Maclaurin sum
formula when turning sums into integrals
\cite{conf,vladb}. We find

\begin{equation}
E_{GS}(N)=\int_{-x_{F}}^{x_{F}}dy\ \epsilon _{0}(y+i\pi
/2+i\omega
/4)\
\rho(y)-\frac{\pi}{6N}\frac{\epsilon^\prime(x_F)}{2\pi\rho(x_F)}.
\end{equation}

This can be expressed in terms of the Fermi velocity
\be
v_F=\frac{\partial\epsilon(x)/\partial x}{\partial p(x)/\partial
x}\biggr|_{x=x_F}=\frac{\epsilon^\prime(x_F)}{2\pi\rho(x_F)}\ ,
\label{vfermi}
\ee
where $p(x)$ is the dressed momentum
\be
p(x)=\int_{-x_F}^{x_F} dx\ \rho(x)\ p_0(x)\ .
\ee
As a result we have that the central charge of the critical
theory is,
as expected, $c=1$.

\item {\underbar{Region~(2):}} For any $v<0$, we have that $\epsilon
_{0}(x+i\pi /2+i\omega /4)>0$ and $\epsilon _{0}(x+i\omega
/4)>0$. As
strings do not play a role, the ground state is the one with no
ledges.

\item {\underbar{Region~(3):}} In this region we need to analyse
the Bethe Ansatz equations (\ref{bae3}). As we already mentioned
we
encounter three different phases by varying $v$. For $v$ close to
zero
the ground state is obtained by filling a Fermi sea of rapidities
$\lambda =x+i\omega /4$ with $x\in [-\pi/2,\pi/2]$.
The ground state energy is given by (\ref{Egs}) with
$x_{F}=\pi/2$,
where the dressed energy $\epsilon (x)$ fulfills
\be
\epsilon(x)
+\frac{1}{2\pi} \int_{-x_{F}}^{x_{F}}dy\
\frac{2\sinh(2\gamma)}{\cosh(2\gamma)-\cos(2x-2y)}\ \epsilon (y)
= \epsilon_0(x) \ ,
\label{dresseden2}
\ee
but does not vanish at the Fermi rapidity $\pi/2$,
{i.e.} $\epsilon (\pm \pi/2)<0$. The bare energy and momentum
that
enter (\ref{Egs}) and (\ref{dresseden2}) are
given by
\begin{eqnarray}
\epsilon _{0}(x) &=&-\log \left| \frac{\cosh (\gamma
+\omega /2)-\cos (2x)}{
\cosh (\gamma -\omega /2)-\cos (2x)}\right| -4v\ , \nonumber \\
\rho _{0}(x) &=&-\frac{1}{2\pi}\frac{dp_0}{dx}=
\frac{1}{\pi }\left( \frac{\sinh (\gamma +\omega
/2)}{\cosh (\gamma +\omega /2)-\cos 2x}
+\frac{\sinh (\gamma -\omega
/2)}{\cosh
(\gamma -\omega /2)-\cos 2x}\right) .
\label{Krho03}
\end{eqnarray}
As $v$ increases in magnitude $x_{F}$ stays at $\pi/2$ until, for
$|v|$
larger than some critical value $v_{c}$, it starts to decrease.
The critical
value is characterized by
\begin{equation}
\epsilon (\pi/2)\bigg|_{v=v_{c}}=0\ .
\end{equation}
For values of $|v|$ slightly larger than $v_{c}$ we are in
Phase~1. Energy
and ledge density are obtained from the same integral equations
as before,
but now $x_{F}<\pi/2$. Finally, as $|v|$ continues to increase it
reaches a
second critical value $v_{c}^{\prime }$ at which a transition to
the
ferroelectric Phase~2 occurs and the ground state becomes
trivial.
\end{itemize}

\begin{figure}[ht]
\begin{center}
\noindent
\epsfxsize=0.6\textwidth
\epsfbox{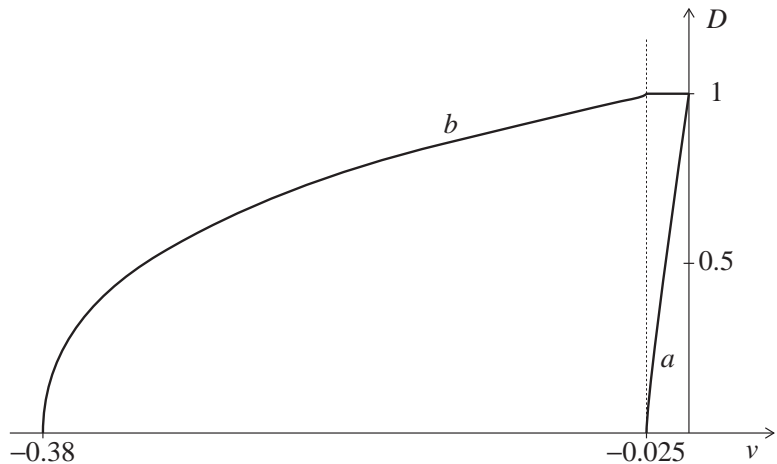}
\end{center}
\caption{\label{fig:ro}
Density as a function of the ledge chemical potential $v$ for two
points in the ($\gamma,\omega$)-plane.}
\end{figure}

In Fig.~\ref{fig:ro} we plot the ledge density as a function $v$
for two
points in the $(\gamma,\omega)$ plane. These points, denoted by
$a$ and $b$,
are chosen such that they lie on phase boundaries for $v=-0.025$
(see Fig.~
\ref{fig:9}). For $v=0$ point $a$ is within the Phase~1 and the
ledge
density is 1. As $|v|$ decreases to $v=-0.025$ the ledge density
decreases
until it becomes zero: there is a transition to Phase~2. For
$v<-0.025$ we
remain in Phase~2 and accordingly $D=0$. Point $b$ is in the
massive Phase~3
for $v=0$ and as $v$ decreases from $0$ the ledge density
remains $1$ until
at $v_c=-0.025$ we enter the Phase~1 and $D$ starts to decrease
until it
reaches zero at $v_c^\prime=-0.38$. For $v<v_c^\prime$ point $b$
belongs to
Phase~2.

\section{Height correlation functions}

\begin{figure}[ht]
\begin{center}
\noindent
\epsfxsize=0.4\textwidth
\epsfbox{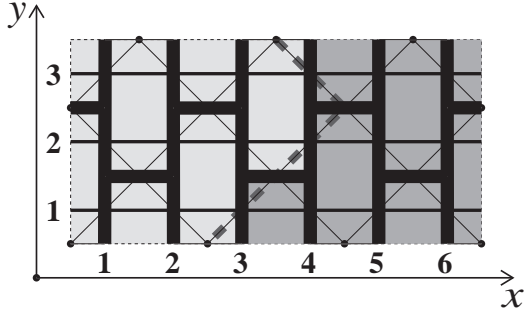}
\end{center}
\caption{\label{fig:6}
Local TLK configuration (for the square geometry).}
\end{figure}

We now define a function $h(x,y)$, which measures the height of
the surface at position $(x,y)$, where the coordinate system is
such
that the transfer direction is parallel to the $y$ axis.
Due to the lattice structure, $h$ is constant on rectangles of
size
$(1,2)$ centred on points $(x+1/2,y-1/2)$, where $x$ and $y$ are
integers with the same parity (see Fig.~\ref{fig:6}). It is thus
sufficient (and indeed most convenient) to measure heights
between
such points only.

Given the way that we defined the geometry of the surface in
terms
of ledges and kinks, which are directly related to the vertex
model,
it is clear that height correlation functions are nonlocal
quantities. We therefore consider height-differences. For any
point
$(x,y)$ where the two coordinates have the same parity, the
occupancy
of the link centred on point $(x,y)$ is related to the height
difference $d_-(x,y)=h(x+1/2,y-1/2)-h(x-1/2,y+1/2)$.
This implies that
$d_-(x,y)=0$ if the link is empty
and $d_-(x,y)=+1$ if it is occupied. If $x$ and $y$ have
different parities then we define
$d_+(x,y)=h(x+1/2,y+1/2)-h(x-1/2,y-1/2)$. Here also
$d_+(x,y)=+1$ if
the link at $(x,y)$ is occupied and $d_+(x,y)=0$ otherwise.

The knowledge of the functions $d_\pm$ is sufficient to
construct the height function $h$ up to an overall additive
constant.

Since we impose periodic boundary conditions (in height
differences),
the expectation values of $d_+$ and $d_-$ are independent of the
position. Moreover, because of invariance under the exchange of
left
and right, we have $\langle d_+\rangle=\langle d_-\rangle=D/2$,
where
$D$ is the ledge density $D=M/N$ (the system size of the system
is taken to be $2N\times 2L$).

The correlation function $\langle d_{\alpha_1}(x_1,y_1)
d_{\alpha_0}(x_0,y_0)\rangle$,
where $\alpha_i=(-1)^{1+x_i+y_i}$ for $i=0,1$,
depends only on $x_1-x_0$ and $y_1-y_0$ because of
two-step translational invariance and left-right and up-down
symmetries. Any height-difference correlation function of the
model
can therefore be obtained from the following connected
correlation
function, which is defined for all integer values of $x_1-x_0$
and
$y_1-y_0$:
\be
C^c(x_1-x_0,y_1-y_0)=\left\langle
(d_{\alpha_1}(x_1,y_1)-D/2)(d_{\alpha_0}(x_0,y_0)-D/2)
\right\rangle .
\ee

Since we have chosen to work with the two-step transfer matrix
$\tau_{dd}$, we need to
distinguish between even and odd values of $y$.
Our convention is that $\tau_{dd}$ connects horizontal lines
with odd values of $y$.

The correlation function for even values $2y$ of the vertical
separation $y_1-y_0$ is given by
\be
C^c(x,2y)=\frac{1}{4}{\rm Trace}\left(
(1+\sigma^z_1-D)\tau_{dd}^y
(1+\sigma^z_{x+1}-D)\tau_{dd}^{L-y}
\right).
\ee
For odd separations in the $y$ direction we find that
\be
C^c(x,2y+1)=\frac{1}{4}{\rm Trace}\left(
U_{2N}^{-1}(1+\sigma^z_1-D)U_{2N}\tau_{dd}^y
(1+\sigma^z_{x}-D)\tau_{dd}^{L-y}
\right).
\ee
The roughness properties of the surface are usually deduced from
the
behaviour of the height correlation function
\be
H^c(x_1-x_0,y_1-y_0)=\left\langle\left(
h(x_1,y_1)-h(x_0,y_0)-(x_1-x_0)D/2
\right)^2\right\rangle .
\ee
This function is related to the connected height-difference
correlation function $C^c$ by
\bea
C^c(x,y)&=&\frac{1}{2}\left[H^c(x+1,y+1)+H^c(x-1,y-1)
-2H^c(x,y)\right]
\nn
&=&\frac{1}{2}\left[H^c(x+1,y-1)+H^c(x-1,y+1)-2H^c(x,y)\right],
\label{ceven}
\eea
if $x$ and $y$ have the same parity and
\be
C^c(x,y)=\frac{1}{2}\left(H^c(x+1,y)+H^c(x-1,y)
-H^c(x,y+1)-H^c(x,y-1)\right)
\label{codd}
\ee
otherwise. $C^c$ is thus a second lattice derivative of $H^c$.

Finally we note that for any state $|M,\alpha\rangle$ in the
$M$-ledge
sector the following sum rule holds:
\be
\left(\sum_{x=1}^{2N}(1+\sigma^z_x-D)\right)|M,\alpha\rangle=0 .
\ee
This implies that for any $y$:
\be
\sum_{x=1}^{2N}C^c(x-x_0,y)=0.
\label{sumrule}
\ee

\section{The massive phase}

In the massive phase the connected link-link correlation function
$C^c(x,y)$ decays exponentially in $x$ and $y$. Exact expressions
are
presently not available although they may in principle be
obtained by
the method of \cite{vladb,fredh}.

It follows that $H^c(x,y)-E_xx-E_yy-F$ (where $E_x$, $E_y$ and
$F$ are
integration constants) also decays exponentially as $x,y\to
+\infty$.

Using the sum rule \r{sumrule} for $x_0=N$ and $y=0$
we can show that $E_x=0$: Inserting \r{ceven} and \r{codd} into
\r{sumrule} we obtain
\be
H^c(N+1,1)-H^c(1-N,1)+H^c(-N,0)-H^c(N,0)=0\ .
\label{asympt}
\ee
Due to left-right symmetry we have $H^c(x,y)=H^c(-x,y)$, which
implies that $H^c(N+1,1)=H^c(N-1,1)$. This exact relationship
is compatible with the large-distance behaviour \r{asympt} only
if $E_x=0$.

This argument extends to any asymptotic formula similar to
\r{asympt}
provided that all the additional terms have a derivative with
respect
to $x$ which vanishes for large values of $x$. As we shall see
below,
this is the case in the massless phase as well.

Returning to the massive phase, if both $E_x$ and $E_y$
vanish, then the surface is smooth, since $H^c$ decays
exponentially to a fixed value $F$. The width
of the interface is of order $\sqrt{F}$.

However, if $E_y>0$, then the height correlations in the $y$
direction
have fluctuations of order $\sqrt{y}$, similar to those of a
one-dimensional string.
We anticipate that $E_y=0$, but we have not been able to prove
that this is the case.

\section{The massless phase}

We now consider the massless phase (Phase~1).
We recall that this phase can be obtained with $b$ and $w$ in
Region~(1)
or Region~(3), in a certain range of values of $v$, which we have
described above.

In both cases the maximal eigenvalue corresponds (in the
thermodynamic
limit) to a filled Fermi sea of Bethe Ansatz rapidities, with a
density function given by \r{rhointeq}, where the kernel and the
bare
density are given by \r{Krho01} or \r{Krho03}.

We use conformal field theory \cite{cft} to relate the
finite-size
spectrum
of the transfer matrix to the long-distance behaviour of
correlation
functions. Using the transfer matrix formalism, it is possible to
determine the spectrum of eigenvalues of $\tau_{dd}$ for
$L\to\infty$,
$N\gg 1$ finite. By conformal covariance, the finite-size spectrum
in the toroidal geometry of the transfer matrix is related to the
power-law
decay of correlation functions in the infinite plane
\cite{cardy}. In
other words, due to conformal symmetry in the critical phase, we
can
calculate the asymptotic behaviour at large distances of
correlation
functions in the infinite system from the finite-size (in $N$)
spectrum of $\tau_{dd}$.

Using standard techniques \cite{conf,vladb}, based on application
of the Euler-Maclaurin sum formula
to taking the thermodynamic limit of
the Bethe Ansatz equations \r{baeq}, we obtain the following
expression for the finite-size spectrum:
\bea
P-P_0&=&2p_Fd+\frac{2\pi}{N}(M^+-M^-+d\Delta M)\\
E-E_0&=&\frac{2\pi v_F}{N}\left(
\left(\frac{\Delta M}{2Z}\right)^2+(Zd)^2
+M^++M^- \right) .
\label{fscorr}
\eea
Here the integers $d$, $\Delta M$ and $M^\pm$ are quantum numbers
characterizing the intermediate states in the spectral
representation
of correlation functions. Given that the largest eigenvalue(s) of
the
transfer matrix can be thought of as (excitations over) a filled
Fermi
sea of rapidities subject to a Pauli principle \cite{kor}, these
quantum numbers can be interpreted as follows:
$d$ is the number of particles backscattered
(``$2p_F$-excitations''),
$\Delta M$ the overall change in the number of particles, and the
$M^\pm$ refer to the creation of particle-hole pairs near
the Fermi points $\pm p_F$.

The other quantities entering \r{fscorr} are the Fermi velocity
$v_F$
defined in \r{vfermi} and the dressed charge $Z=Z(x_F)$, which is
defined in terms of the integral equation
\be
Z(x)+\frac{1}{2\pi}\int_{-x_F}^{x_F}dy\ K(x-y)\ Z(y) =1\ ,
\ee
where the kernel $K(x)$ is defined in \r{Krho01} or
\r{dresseden2}.

\begin{figure}[ht]
\begin{center}
\noindent
\epsfysize=0.4\textwidth
\epsfbox{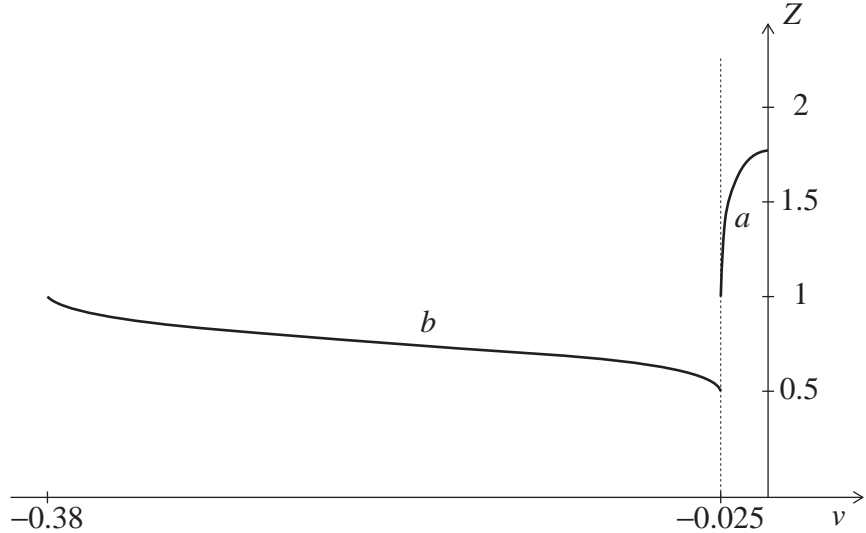}
\end{center}
\caption{\label{fig:Z}
Dressed charge as a function of ledge chemical potential for two
points in the ($\gamma,\omega$)-plane.}
\end{figure}

\begin{figure}[ht]
\begin{center}
\noindent
\epsfysize=0.3\textwidth
\epsfbox{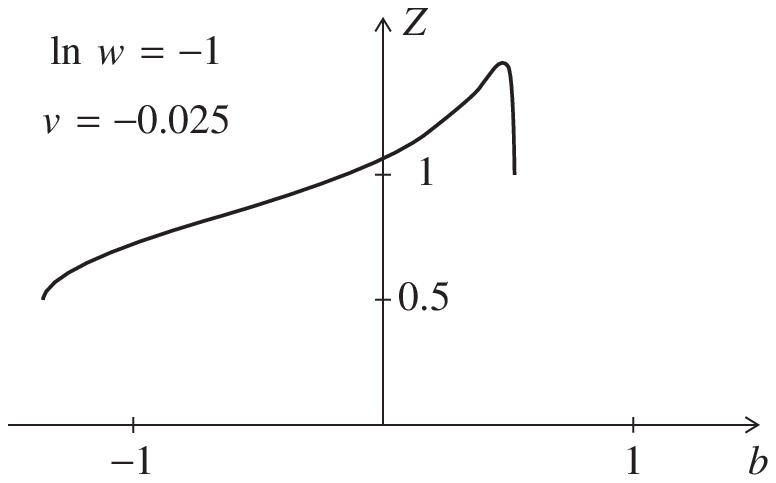}
\hfill
\epsfysize=0.3\textwidth
\epsfbox{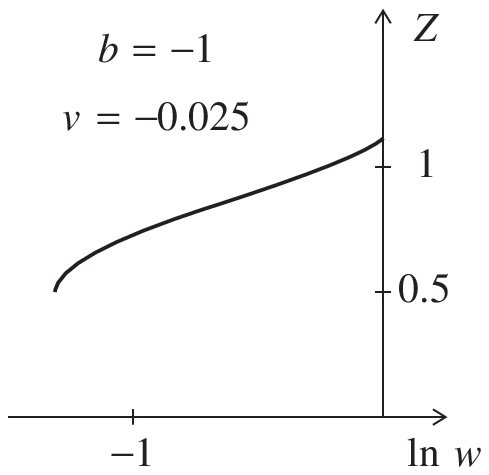}
\end{center}
\caption{\label{fig:Zbc}
(a) Dressed charge as a function of ledge interaction for fixed
values
of ledge stiffness $w$ and chemical potential $v$ (b) Dressed
charge
as a function of $\ln w$ for fixed $b$ and $v$.}
\end{figure}

Since the correlation functions in which we are interested
contain only operators which conserve the number of particles,
we can restrict ourselves to $\Delta M=0$.
We then obtain the following form for the asymptotics of the
height-difference correlation function
\be
C^c(x,y)=\frac{A}{x^2+v_F^2y^2}
+\frac{B\cos(2p_Fx)}{(x^2+v_F^2y^2)^{Z^2}} .
\label{ccxy}
\ee
It is useful to recast \r{ccxy} in a slightly different form.
We consider a line in the $(v_F\sin\theta,\cos\theta)$ direction
($\theta=0$
is vertical, $\theta=\pi/2$ is horizontal) and define a height
function
$h_\theta(r)=h(rv_F\sin\theta,r\cos\theta)$. In the scaling limit
height-differences become derivatives. Thus, we expect the
lattice
equivalent of $h_\theta^\prime=dh_\theta(r)/dr$ to be
$\frac{\cos\theta+v_F\sin\theta}{2}d_+
+\frac{\cos\theta-v_F\sin\theta}{2}d_-$, which in turn implies
the
following form for the connected correlation function in radial
coordinates
\be
C^c_\theta(r_2-r_1)=\left\langle
h_\theta^\prime(r_1)h_\theta^\prime(r_2)\right\rangle=
A_\theta
|r_2-r_1|^{-2}+B_\theta\cos(2p_Fv_F|r_2-r_1|\sin\theta)r^{-2Z^2}
\ .
\ee
Here $A_\theta$ and $B_\theta$ are quadratic forms in
$\cos\theta\pm v_F\sin\theta$, i.e. affine functions of
$\cos 2\theta$ and $\sin 2\theta$.

$C^c_\theta(r)$ is related to the height correlation function
\be
H^c_\theta(r_2-r_1)=\left\langle\left(h_\theta(r_1)-h_\theta(r_2)
-\frac{D}{2}(r_1-r_2)v_F\sin\theta\right)^2
\right\rangle
\ee
by
\be
C^c_\theta(r)=-\frac{d^2H^c_\theta(r)}{dr^2}
\ee
therefore the asymptotic behaviour of $H$ is
\be
H^c_\theta(r)=A_\theta\ln
r+\frac{B_\theta}{4p_F^2v_F^2\sin^2\theta}
\frac{\cos(2p_Fv_Fr\sin\theta)}{r^{2Z^2}}+E_\theta r+F_\theta
\ee
We believe that $E_\theta=0$ for any value of $\theta$.
We have already seen that $E_{\pi/2}=0$.

As $\theta\to 0$ the amplitude
of the $r^{-2Z^2}$ term is expected
to diverge. We cannot prove this
since $B_\theta$ might be proportional
to $(1-\cos 2\theta)$, but we have
no reason to believe
that this is the case.

\section{Discussion}

The results of \cite{dba} were mainly of a thermodynamic
character.  In the half-filling case, corresponding to
a tilt angle of $\pi/4$, there is a phase transition
at $e^b=(1-w)^2$ for the case of repulsive interactions
$b<0$.  This is of Kosterlitz-Thouless, or equivalently,
Lieb-F type.  For tilt angles $0<\theta<\theta_f$
which differ from $\pi/4$, there is no
thermodynamic singularity.  For $\theta=\pi/4$
and small enough temperature, a Peierls proof along
the lines of Brascamp et al. \cite{brascamp}
shows that the fluctuations
in height about the $\theta=\pi/4$ surface are of order
one, whereas in the free-Fermi case $e^b=1-w^2$
(see appendix B), there are logarithmic height fluctuations,
as in the BCSOS model, which also has a free-Fermion
point.

The interesting question is what happens for $\theta\neq\pi/4$
at a level of correlation functions.  We show that for
$\theta\neq\pi/4$ and no binding of ledges, the excitation
spectrum is gapless.  This gives a power-law decay of
height step-step correlation functions by using conformal
arguments.  In addition to the $r^{-2}$ term which leads
to logarithmic height fluctuations, we find
an oscillatory term which decays algebraicly with a
nonuniversal exponent, but whose contribution to height
fluctuations is subdominant.

At $\theta=\pi/4$ and below the transition temperature,
the transfer-matrix spectrum has a gap, giving
exponential decay of correlations of local objects.

There is another phase transition which occurs for any angle
$0<\theta<\theta_f$ at the solution of $e^b=(1+w)^2$.
This has zero limiting entropy per unit basal area below
the transition temperature $T_c$, and a specific heat
which diverges as $(T-T_c)^{-1/2}$ from above.
In this high-temperature phase, we have gapless excitations
as before, and therefore a rough surface.  This phase
transition may be thought of as of pinning-depinning
type.  The low-temperature phase has a complete collapse
of the ledges in a close-packed strip with axis in the
Euclidean time direction.  We have not analysed the
correlation functions in this case.  We must point out
that the results in this part of the phase diagram
are almost certainly critically dependent on the nature
of the short-range interactions.

In appendix E, the interaction between two ledges
separated by mean distance $l$ is
found to have the same $l^{-2}$ behaviour
as the law
which would be obtained from elastic
theory or from the interaction between
electrostatic dipoles located along
the ledges, which may be approximated
by pairwise interactions between neighbouring
edges.  Our model gives a consistent
treatment of that part of the elastic interaction
which comes from deformation of the surface
via entropic interactions.  This part tends
to zero as the temperature $T\to 0$.  But the dipole
term does not vanish in the same limit and thus,
when present, should dominate the low-temperature
behaviour.  Thus we believe that the model considered by
Villain et al. \cite{villain} for
sections vicinal to Cu(111) is fundamentally
different;
our model only shows roughening at the unique
half-filling angle of tilt, thermodynamically
speaking, and the microscopic picture strongly
supports this view.
For less-than-half filling, there is extensive degeneracy
because of the form of the short-range interaction
in our basic model.  One might then ask why the
$1/l^2$ entropic repulsion which comes in at
the coarse-grained level does not break this
degeneracy, for any $T>0$ (at zero temperature
this interaction vanishes), and thereby produce a
phase transition.  Clearly a better understanding
of this is needed.

The recent work on vicinal surfaces of $^4
{\rm He}$ crystals
\cite{4he} suggests that there may be a critical
tilt angle of the vicinal section, below which neighbouring
ledges are rather straight and thus do not overlap, but
above which the ledges wander sufficiently to come into
close contact.  In our model, such a critical angle
does not appear in the thermodynamics.
However, in order to address the questions raised
by the experiments \cite{4he}, it may be
necessary to study the behaviour of ledges
at a microscopic level, and possibly also
to take into account defects and boundaries.
Examples are known \cite{ntt} of pair functions
near surfaces with weakened bonds where there is a transition
behaviour in the asymptotics, but, on applying a fluctuation
sum formula to get a derivative of the free energy,
no thermodynamic singularity is seen.  Thus
caution is advisable in drawing conclusions from 
the thermodynamics alone.

\section{Acknowledgments}

We are grateful to B. Duplantier, T.~L. Einstein,
M. Gaudin, J.~M. Luck, A.~E. Malevanets
and N. Mousseau for interesting discussions.

We acknowledge financial support from EPSRC
under grant number GR/K97783.

\appendix

\section{A commuting family of transfer matrices for the TLK
model}

In this appendix we construct an embedding of the TLK model into
the
framework of the Quantum Inverse Scattering Method \cite{vladb}.
We then
show how to recover the coordinate Bethe Ansatz solution of
\cite{dba}. The
vertices in the diagonal-to-diagonal formulation of the TLK
model are given
by (\ref{weights}). We define a diagonal-to-diagonal transfer
matrix as
follows
\begin{equation}
U_{2N}=R^{\beta_2\beta_3}_{\alpha_1\alpha_2}
R^{\beta_4\beta_5}_{\alpha_3\alpha_4}\ldots
R^{\beta_{2N}\beta_1}_{\alpha_{2N-1}\alpha_{2N}}\ .
\end{equation}
We note that the square of this transfer matrix commutes with
the square of
the translation operator $\tau_R$
\begin{equation}
[U_{2N}^2,\tau_R^2]=0\ .
\end{equation}
The transfer matrix $\tau_{dd}$ studied in \cite{dba} is
expressed as
\begin{equation}
\tau_{dd}=\tau_R^2\ U_{2N}^2\ ,
\end{equation}
and the partition function of the TLK model is given by
\begin{equation}
Z={\rm Trace}\left( \tau_{dd}\right)^L\ .
\end{equation}
Correlation functions can also be expressed in terms of
$\tau_{dd}$ so that
the problem of solving the TLK model reduces to the one of
finding
eigenvalues and eigenvectors of $\tau_{dd}$.

Following \cite{diag} we can reformulate
the problem in terms of an
inhomogeneous row-to-row transfer matrix of a 6-vertex model in
a vertical
electric field \cite{a6v}. Starting point is the usual XXZ
(symmetric
6-vertex) $L$-operator \cite{vladb}
\begin{equation}
L_n(\lambda)=\left(
\begin{array}{cc}
\frac{\sinh(\lambda+i\frac{\gamma}{2}\sigma_n^z)}{i\sin\gamma} &
\sigma_n^-
\\[8pt]
\sigma_n^+ &
\frac{\sinh(\lambda-i\frac{\gamma}{2}\sigma_n^z)}{i\sin\gamma}
\end{array}
\right).
\end{equation}
Here $\Gamma$ is a free parameter and $\sigma^\alpha$ are the
usual Pauli
matrices. Note that there exists a so-called shift point at
$\lambda=i\frac{
\gamma}{2}$ (at which the L-operator reduces to the permutation
operator), i.e.
\begin{equation}
L_n(i\frac{\gamma}{2})= \left(
\begin{array}{cc}
\frac{1}{2}(1+\sigma^z_n) & \sigma_n^- \\[8pt]
\sigma_n^+ & \frac{1}{2}(1-\sigma^z_n)
\end{array}
\right). \label{shift}
\end{equation}
The L-operator of the model with a vertical field is then simply
\begin{equation}
{\cal L}_n(\lambda)=\exp(\frac{v}{2}\sigma^z_n)L_n(\lambda)
\exp(\frac{v}{2}
\sigma^z_n)\ .
\end{equation}
Consider now an {\sl inhomogeneous} row-to-row transfer matrix
of such an
asymmetric 6-vertex model on a lattice with $2N$ sites
\begin{eqnarray}
{\cal T}(\mu)&=&{\rm Trace}\ {\cal
L}_{2N}(\mu-i\frac{\omega}{2}){\cal L}
_{2N-1}(\mu) {\cal L}_{2N-2}(\mu-i\frac{\omega}{2}){\cal L}
_{2N-3}(\mu)\ldots {\cal L}_2(\mu-i\frac{\omega}{2}){\cal
L}_{1}(\mu)
\nonumber \\
&\equiv& {\rm Trace}\
\exp(\frac{v}{2}\sum_n\sigma^z_n)T(\mu)\exp(\frac{v}{2}
\sum_n\sigma^z_ n) \ ,
\end{eqnarray}
where $T(\mu)$ is the symmetric inhomogeneous 6-vertex transfer
matrix
constructed from the L-operators $L_n(\lambda)$. Here the
inhomogeneity $\Omega$ is a second free parameter. Choosing
$\mu=i\frac{\gamma}{2}$ and
using (\ref{shift}) we find
\begin{equation}
{\cal T}(i\frac{\gamma}{2})= \left(\widehat{{\cal L}}_{2N}
(i\frac{\gamma}{2}
-i\frac{\omega}{2})\right)^{\alpha_1\beta_{2N-1}}_{\alpha_{2N}\
b_{2N}}
\left(\widehat{{\cal L}}_{2N-2}
(i\frac{\gamma}{2}-i\frac{\omega}{2}
)\right)^{\alpha_{2N-1}\beta_{2N-3}}_{
\alpha_{2N-2}\beta_{2N-2}} \ldots
\left(\widehat{{\cal L}}_{2}(i\frac{\gamma}{2}-i\frac{\omega}{2}
)\right)^{\alpha_3\beta_1}_{\alpha_2\beta _2}\ ,
\end{equation}
where
\begin{equation}
\widehat{{\cal L}}_{2N}(i\frac{\gamma-\omega}{2})= \left(
\begin{array}{cccc}
e^{2v}\frac{\sin({\gamma}-\frac{\omega}{2})}{\sin {\gamma}} & &
& \\[8pt]
& -\frac{\sin\frac{\omega}{2}}{\sin {\gamma}} & 1 & \\[8pt]
& 1 & -\frac{\sin\frac{\omega}{2}}{\sin {\gamma}} & \\[8pt]
& & & e^{-2v}\frac{\sin({\gamma}-\frac{\omega}{2})}{\sin
{\gamma}}
\end{array}
\right).
\end{equation}
It follows that the inhomogeneous row-to-row transfer matrix is
identical to
the diagonal-to-diagonal transfer matrix if we make the following
identifications
\begin{equation}
e^{-2v}\frac{\sin({\gamma}-\frac{\omega}{2})}{\sin
{\gamma}}=e^{-a}\ ,\quad -
\frac{\sin\frac{\omega}{2}}{\sin {\gamma}}=w\ ,\quad
\left(\frac{\sin({\gamma
}-\frac{\omega}{2})}{\sin {\gamma}}\right)^2=e^{b}\ .
\end{equation}
Note that $4v=b+2a$. Consider now an eigenstate of $T(\mu)$ (the
transfer
matrix of the inhomogeneous, symmetric 6-vertex model) with
eigenvalue $\nu(\mu)$
\begin{equation}
T(\mu)|\Lambda\rangle=\nu(\mu)|\Lambda\rangle\ .
\end{equation}
Then, because the $z$-component of total spin is a good quantum
number, we
have
\begin{equation}
{\cal T}(\mu)|\Lambda\rangle=
e^{v(N_\uparrow-N_\downarrow)}\nu(\mu)|\Lambda\rangle\ ,
\end{equation}
where $N_{\uparrow,\downarrow}$ are the total numbers of up
up/down spins in
the state $|\Lambda\rangle$. This means that we can obtain a
complete set of
eigenstates of ${\cal T}(\mu)$ from a complete set of
eigenstates of $T(\mu)$.
The eigenvalues of $T(\mu)$ are given by (we choose the state
with all
spins down as the reference state)
\begin{eqnarray}
\nu(\mu)&=&\left(\frac{\sinh(\mu-i\frac{\omega}{2}-i\frac{\gamma}
{2})
\sinh(\mu-i\frac{\gamma}{2})}{-\sin^2\gamma}\right)^N\prod_{j=1}^
M \frac{
\sinh(\mu-\lambda_j+i\gamma)}{\sinh(\mu-\lambda_j)} \nonumber \\
&+&\left(\frac{\sinh(\mu-i\frac{\omega}{2}+i\frac{\gamma}{2})
\sinh(\mu+i
\frac{\gamma}{2})}{-\sin^2\gamma}\right)^N\prod_{j=1}^M \frac{
\sinh(\mu-\lambda_j-i\gamma)}{\sinh(\mu-\lambda_j)}\ , \label{EV}
\end{eqnarray}
where $M=N_\up$ and where the spectral parameters $\lambda_j$
are solutions
of the Bethe Ansatz equations
\begin{equation}
\left( \frac{\sinh\left(\lambda_k+i\frac{\gamma}{2}\right)} {
\sinh\left(\lambda_k-i\frac{\gamma}{2}\right)}\
\frac{\sinh\left(\lambda_k-i
\frac{\omega-\gamma}{2}\right)}
{\sinh\left(\lambda_k-i\frac{\omega+\gamma}{2
}\right)}\right)^N= -\prod_{j=1}^M
\frac{\sinh\left(\lambda_k-\lambda_j+i{
\gamma}\right)}
{\sinh\left(\lambda_k-\lambda_j-i{\gamma}\right)}\ ,\
k=1\ldots M\ .
\label{bae}
\end{equation}
We thus find that the eigenvalues of $U_{2N}={\cal
T}(\frac{i\gamma}{2})$ are
given by
\begin{equation}
\Lambda(\{\lambda_j\})= e^{a(M-N)}
\left(\frac{\sin(\gamma-\frac{\omega}{2})
}{\sin\gamma}\right)^{M}
\prod_{j=1}^M\frac{\sinh\left(\lambda_j+i\frac{
\gamma}{2}\right)}
{\sinh\left(\lambda_j-i\frac{\gamma}{2}\right)}\ .
\label{ev2}
\end{equation}
In order to make contact with the results obtained in \cite{dba}
we need to
consider the transfer matrix $\tau_{dd}=\tau_R^2U_{2N}^2$. From
the standard
construction of \cite{vladb} it follows that the eigenvalues of
the operator
of translation by two sites $\tau_R^2$ are given by
\begin{equation}
\exp\left(ip(\{\lambda_j\})\right)=\prod_{k=1}^M
\frac{\sinh\left(\lambda_k-i
\frac{\gamma}{2}\right)}
{\sinh\left(\lambda_k+i\frac{\gamma}{2}\right)}\
\frac{\sinh\left(\lambda_k-i\frac{\omega+\gamma}{2}\right)} {
\sinh\left(\lambda_k-i\frac{\omega-\gamma}{2}\right)}\ .
\label{trans}
\end{equation}
This then implies that the eigenvalues $\Lambda_{dd}$ of
$\tau_{dd}$ are
\begin{equation}
\Lambda_{dd}(\{\lambda_j\})=e^{2a(M-N)}
\left(\frac{\sin(\gamma-\frac{\omega
}{2})}{\sin\gamma}\right)^{2M}
\prod_{j=1}^M\frac{\sinh\left(\lambda_j+i
\frac{\gamma}{2}\right)}
{\sinh\left(\lambda_j-i\frac{\gamma}{2}\right)}
\frac{\sinh\left(\lambda_j-i\frac{\gamma+\omega}{2}\right)} {
\sinh\left(\lambda_j+i\frac{\gamma-\omega}{2}\right)}\ .
\label{ev3}
\end{equation}
Equations (\ref{bae}) and (\ref{ev3}) are equivalent to (11) and
(9) of \cite{dba}
if we set $a=0$ and substitute the function $s(k)$ of
\cite{dba} by
\begin{equation}
s(k)=\frac{\sinh (\lambda-i\omega/2+i\gamma/2)}{\sinh (\lambda
-i\gamma/2)}=
\frac{\exp(2\lambda+i\gamma-i\omega/2)-\exp(i\omega/2)}
{\exp(2\lambda)-
\exp(i\gamma)}\ .
\end{equation}

\section{Free-fermion cases}

\label{free}
For $\gamma=\pi/2 $ the two-particle scattering
phase shifts on the right hand side of (\ref{baeq}) reduce to
$-1$:
the particles are
free fermions. This corresponds to the following constraint on
the weights:
$e^b=1-w^2$. The model is physical only if $e^b$ and $w$ are
real and
positive. This implies that
$b$ is negative: the interaction between
ledges is repulsive.

The Bethe Ansatz equations (\ref{bae}) become
\begin{equation}
\left(\frac{\sinh (\lambda _{k}+i\gamma/2)\sinh (\lambda _{k}
-i\omega/2+i\gamma/2)}{\sinh (\lambda _{k}-i\gamma/2)\sinh
(\lambda
_{k}-i\omega/2-i\gamma/2)}\right) ^{N} =(-)^{M-1}
\end{equation}

Clearly the rapidities $\lambda_k$ are constrained to lie on the
lines $\Im
m\lambda =i\omega/4$ and $\Im m\lambda =i\omega/4+i\pi/2$, which
we denote
by ${\cal C}_1$ and ${\cal C}_2$ respectively. The state
corresponding to
the largest eigenvalue of the transfer matrix $\tau_{dd}$ is
then obtained
by filling a Fermi sea of (occupied) rapidities on one of these
lines. We
distinguish two cases:

\begin{itemize}
\item {} \underbar{$w\coth 2v<-1$}:

In this case there is a Fermi sea on ${\cal C}_{2}$ with Fermi
level given
by $\cosh 2\lambda _{F}=-w\coth 2v$. The line ${\cal C}_{1}$ is
empty. This
is a critical phase with power-law correlations.

\item {} \underbar{$-1<w\coth 2v<0$}:

In this case the reference state itself (i.e. the state
with all spins down) corresponds to the largest eigenvalue: both
lines ${\cal C}_{j}$ are empty. This phase is in general massive
i.e. the next-largest eigenvalue of $\tau_{dd}$ is
separated from the largest one by a gap. Accordingly correlations
are characterized by an exponential decay.
\end{itemize}

Note that for $v>0$ the state with all spins up can be used as
reference state; the distribution of rapidities is then given by
the same rule as above but with $v$ changed to $-v$. (We do
not consider this case any further.)

For any $0<w<1$ all regimes can be reached by varying $a$. For
$v=0$, i.e. $a=-b/2$, the number of ledges in the ground
state is $M=N$. In order to study correlation functions it is
convenient to switch to the ``coordinate-space'' notation of
\cite{dba}. We define a ``particle momentum'' $k$ and two
functions $s_\pm(k)$ by
\begin{eqnarray}
e^{ik}&=&\frac{\sinh (\lambda+i\gamma/2)\sinh (\lambda-i\omega/2
+i\gamma/2)
}{\sinh (\lambda-i\gamma/2)\sinh (\lambda-i\omega/2
-i\gamma/2)}, \nn
s_{\pm }(k)&=&\frac{w(e^{ik}-1)}{2}\pm e^{ik/2}\sqrt{1-w^{2}\sin
^{2} k/2 } .
\end{eqnarray}

We then construct fermionic creation and annihilation operators
by means of a Jordan-Wigner transformation on the spin variables
defining the vertex model
\begin{equation}
c_{j}^{+}=(-\sigma_{1}^{z})(-\sigma_{2}^{z})\dots(-\sigma_{j-1}^{
z} )\frac{
\sigma_{j}^{x}+i\sigma_{j}^{y}}{2}.
\end{equation}
In the momentum-space representation there are two ``bands'' of
fermions
with momentum $k$ because rapidities on ${\cal C}_1$ and ${\cal
C}_2$
with equal real parts correspond to the same value of $k$.
Fermion
annihilation and creation operators are of the form
\begin{eqnarray}
C_\pm(k)&=&\sqrt{\frac{2}{N} }\frac{1}{s_{\mp }(k)-s_{\pm }(k)
}\left(
\sum\limits_{j=1}^{N}e^{-ikj} s_{\mp }(k) c_{2j}
-\sum\limits_{j=0}^{N-1}e^{-ikj} c_{2j+1}\right), \nonumber \\
C_\pm^\dagger(k)&=&\frac{1}{\sqrt{2N}}\left(\sum\limits_{j=1}^{N}
e^{ikj}
c_{2j}^{+}+\sum\limits_{j=0}^{N-1}e^{ikj}s_{\pm}(k)c_{2j+1}^{+}
\right).
\end{eqnarray}
The inverse relation between momentum-space and position-space
operators is
\begin{eqnarray}
c_{2j}=\frac{1}{\sqrt{2N} }\sum_{k,\alpha} e^{ikj} C_{k,\alpha }
&\qquad &
c_{2j+1}=\frac{1}{\sqrt{2N} }\sum_{k,\alpha} s_{\alpha }(k)
e^{ikj}
C_{k,\alpha } \\
c_{2j}^{+}=\sqrt{\frac{2}{N} }\sum_{k,\alpha}\frac{s_{-\alpha
}(k) }{
s_{-\alpha }(k)-s_{\alpha }(k) } e^{-ikj} C_{k,\alpha
}^{\dagger} &\qquad &
c_{2j+1}^{+}=\sqrt{\frac{2}{N} }\sum_{k,\alpha} \frac{e^{-ikj}
}{s_{\alpha
}(k)-s_{-\alpha }(k) } C_{k,\alpha }^{\dagger}\ .
\end{eqnarray}
The vacuum state of the fermionic Fock space is defined by
$C_{k,\pm}|\Omega\rangle=0$. Eigenstates of the transfer matrix
are obtained
by acting with creation operators $C^\dagger_{k,\pm}$ on the
vacuum
\begin{equation}
|k_1\ldots k_n|\alpha_1\ldots \alpha_n\rangle= \prod_{j=1}^n
C^\dagger_{k_j,\alpha_j}|\Omega\rangle\ .
\end{equation}
The corresponding eigenvalue of the transfer matrix is
\begin{eqnarray}
&&\tau_{dd}|k_1\ldots k_n|\alpha_1\ldots \alpha_n\rangle=
\prod_{j=1}^n
e^{2a}[1+w\ s_{\alpha_j}(k_j)][1+w/s_{\alpha_j}(k_j)] \
|k_1\ldots
k_n|\alpha_1\ldots \alpha_n\rangle\ .
\end{eqnarray}
In terms of the $k,\alpha$ variables the maximal eigenvalue of
the transfer matrix corresponds to the state $|0\rangle$ defined
by filling a Fermi sea in the $+$ band between the Fermi points
$-k_F$
and $k_F$, i.e.
\begin{equation}
|0\rangle = \prod_{|k|\leq k_F} C^\dagger_+(k)|\Omega\rangle \ .
\end{equation}
Let us now turn to the calculation of correlation functions.
using
standard free-fermion methods we can derive integral
representations
for the correlation functions of height differences. However, we
shall
see that the results are much more complicated than for the case
of the BCSOS model \cite{bcsos,suth}.

\subsection{ Correlators of height-differences in $y$-direction}
The height difference in $y$-direction is given by
$d_y(x,2y)=h(x,2y+1)-h(x,2y-1)$, where $x$ is a half-odd-integer.
The
corresponding operator for $x=3/2$ is given by
\begin{eqnarray}
(\delta_y(3/2))^{\alpha_1\ldots\alpha_{2N}}
_{\alpha^\prime_1\ldots\alpha^\prime_{2N}}
&=&A^{\alpha_1\alpha_2}_{\alpha^\prime_1\alpha^\prime_2}
\prod_{j=3}^{2N}\delta^{\alpha_j}_{\alpha^\prime_j},
\label{defdeltay}
\end{eqnarray}
where the nonzero elements of $A$ are
\begin{eqnarray}
A^{12}_{21}=\frac{w}{1-w^{2}},\qquad
A^{21}_{12}=-\frac{w}{1-w^{2}},\qquad
A^{12}_{12}=\frac{w^2}{1-w^{2}},\qquad
A^{21}_{21}=-\frac{w^2}{1-w^{2}}.
\end{eqnarray}
The correlation function of the height differences (for even
separations) is then given by
\bea
C_{yy}(2x,2y)&=&\langle
d_y(2x_0-1/2,2y_0-1)d_y(2x_0-1/2+2x,2y_0-1+2y) \rangle\nn
&=&{\rm Trace}
\left(\tau_{dd}^{L-y}\delta_y(3/2)\tau_{dd}^y\tau_R^{2x}\delta_y(3/2)
\tau_R^{-2x}\right). \label{dycorr}
\eea
We note that this formula is valid only for $y>0$.
formula
The operator (\ref{defdeltay}) can be expressed in terms of the
fermion operators as
\begin{eqnarray}
\delta_y(3/2)
&=&\frac{1}{N}\sum_{k_{1},k_{2},\alpha_{1},\alpha_{2}}
a_{\alpha_1,\alpha_2}(k_{1},k_{2})\ C_{\alpha_2}^{\dagger}(k_2)\
C_{\alpha_1}(k_1)\ ,
\end{eqnarray}
where
\begin{equation}
a_{\alpha_1,\alpha_2}(k_1,k_2)= \frac{w\
s_{\alpha_1}(k_1)+we^{ik_1-ik_2}s_{\alpha_2}(k_2)+
e^{ik_1}+s_{\alpha_1}(k_1)s_{\alpha_2}(k_2)} {(w-w^{-1})(s_{
\alpha_2}(k_2)-s_{-\alpha_2}(k_2))}\ .
\end{equation}
This yields the following representation for the correlation
function of height-differences
\be
C_{yy}(2x,2y)=
\frac{1}{N^{2}}\sum_{{\rm hole}(k_{h},+),{\rm
particle}(k_p,\alpha)}
a_{+,\alpha}(k_{h},k_{p})\ a_{\alpha, +}(k_{p},k_{h})\
\left(\Lambda_\alpha(k_p)/\Lambda_+(k_h)\right)^{y}
e^{ix(k_p-k_h)}. \label{ff:cyy}
\ee
Here the sum extends over all holes with momentum $k_h$ in the
Fermi sea in
the $+$ band ($|k_h|\leq k_F$) and all particles in either the
$+$ band ($\pi\geq |k_p|> k_F$, $\alpha=+$) or
the $-$ band ($|k_p|\leq
\pi$, $\alpha=-$) and
\begin{equation}
\Lambda_\alpha(k) = e^{2a}[1+w\
s_{\alpha}(k)][1+w/s_{\alpha}(k)]\ .
\end{equation}

In the thermodynamic limit we obtain the following integral
representation
\begin{eqnarray}
C_{yy}(2x,2y) &=&\int_{-k_{F}}^{k_{F}}dk_{h}\left\{
\int_{k_{F}}^{\pi}
+\int_{-\pi }^{-k_{F}}dk_{p}\right\} a_{+,+}(k_{h},k_{p})\
a_{+,+}(k_{p},k_{h})\nn
&&\left( \frac{w\cos
(k_{p}/2)+\sqrt{1-w^{2}\sin
^{2}(k_{p}/2)}}{w\cos (k_{h}/2)+\sqrt{1-w^{2}\sin
^{2}(k_{h}/2)}}\right)
^{2y}e^{ix(k_p-k_h)} \nonumber \\
&+&\int_{-k_{F}}^{k_{F}}dk_{h}\int_{-\pi }^{\pi }dk_{p}\
a_{+,-}(k_{h},k_{p})\ a_{-,+}(k_{p},k_{h})\nn
&&\left( \frac{w\cos
(k_{p}/2)-\sqrt{
1-w^{2}\sin ^{2}(k_{p}/2)}}{w\cos (k_{h}/2)-\sqrt{1-w^{2}\sin
^{2}(k_{h}/2)}}
\right) ^{2y}e^{ix(k_p-k_h)}\ . \label{ff:corr}
\end{eqnarray}
We have not managed to greatly simplify (\ref{ff:corr}).
However, it is straightforward to extract the asymptotic
behaviour for
$z\to \infty $ by expanding the integrand around the maxima at
$k_{h},k_{p}=\pm k_{F}$.

\subsection{Correlators of height-differences in $x$-direction}

The height difference in $x$-direction
$d_{x}(1,2y-1)=h(3/2,2y-1)-h(1/2,2y-1)$ is measured by the
operator
\begin{eqnarray}
\delta_x(1)&=&c_{1}^{-}c_{1}^{+}
=\frac{1}{N}\sum_{k_{1},k_{2},\alpha
_{1},\alpha _{2}}\frac{s_{\alpha
_{1}}(k_{1})}{s_{\alpha _{2}}(k_{2})-s_{-\alpha
_{2}}(k_{2})}C_{\alpha
_{2}}^{\dagger }(k_{2})\ C_{\alpha _{1}}(k_{1})\ .
\end{eqnarray}
The corresponding correlation function of height differences (for
even separations) has the following integral representation
\begin{eqnarray}
C_{xx}(2x,2y) &=&\int_{-k_{F}}^{k_{F}}dk_{h}\left\{
\int_{k_{F}}^{\pi
}+\int_{-\pi }^{-k_{F}}dk_{p}\right\}
\frac{s_+(k_h)}{s_+(k_p)-s_-(k_{p})}
\frac{s_+(k_p)}{s_+(k_h)-s_-(k_{h})}
\nn
&&\left( \frac{w\cos (k_{p}/2)+\sqrt{1-w^{2}\sin
^{2}(k_{p}/2)}}{w\cos (k_{h}/2)+\sqrt{1-w^{2}\sin
^{2}(k_{h}/2)}}\right)
^{2y}e^{ix(k_p-k_h)} \nonumber \\
&+&\int_{-k_{F}}^{k_{F}}dk_{h}\int_{-\pi }^{\pi }dk_{p}\
\frac{s_+(k_h)}{s_-(k_p)-s_+(k_{p})}
\frac{s_-(k_p)}{s_+(k_h)-s_-(k_{h})}\nn
&&\left( \frac{w\cos (k_{p}/2)-\sqrt{
1-w^{2}\sin ^{2}(k_{p}/2)}}{w\cos (k_{h}/2)-\sqrt{1-w^{2}\sin
^{2}(k_{h}/2)}}
\right) ^{2y}e^{ix(k_p-k_h)}\ .
\end{eqnarray}

\subsection{Height-correlations along the $y$-direction}
Above we have derived integral representations for the (local)
correlation functions of height-differences. In the remainder of
this
appendix we consider height-correlations along the transfer
direction.
The height correlation function
$H(0,y)=\langle(h(3/2,2y)-h(3/2,0))^2\rangle$ is related to
$C_{yy}(0,y)$ by
\begin{equation}
H(0,2y+2)+H(0,2y-2)-2H(0,2y)=2C_{yy}(0,2y),
\label{hc}
\end{equation}
where $H(0,0)=0$ and $H(0,2)=C_{yy}(0,0)$. Summing \r{hc} we find
that
\begin{equation}
C_{yy}(0,0)+2\sum_{z=1}^{y-1}C_{yy}(0,2z)=H(0,2y)-H(0,2y-2).
\end{equation}
The algebraic decay $C_{yy}(0,2y)\propto \frac{A}{y^2}$
of the height-difference correlation functions
implies that the surface is rough
\begin{equation}
H(0,2y)\propto F + E_y\ y + A^\prime\ \ln y +\ldots\ ,
\end{equation}
where in the thermodynamic limit
\begin{equation}
E_y=C_{yy}(0,0)+2\sum_{y=1}^{\infty}C_{yy}(0,2y)\ . \label{ey}
\end{equation}
On physical grounds we expect that $E_y=0$ as we now argue.
Let us define a variable $D_y=\sum_{y=1}^{N}d_y(x,2y)$ in our
finite
$2N\times 2L$-toroidal geometry. $D_y$ is independent of $x$
because
$d_x(x+1/2,y)=h(x+1,y)-h(x,y)$ is subject to periodic boundary
conditions. $E_y$ and $D_y$ are related by
\begin{equation}
E_y=\lim_{N,L\rightarrow\infty}\left\langle d_y
(x,y)D_y\right\rangle=\lim_{N,L\rightarrow\infty}
\frac{\left\langle
D_y^{2}\right\rangle}{L}.
\end{equation}
The average tilt of ledges in the $xy$-plane is $s=(ND_y)/(LM)$:
for
instance if $D_y=0$ the ledges are on average parallel to the
transfer
direction ($s=0 $). Phenomenologically we expect that ledges
have a
stiffness $\Sigma$, so that the total energy cost of a nonzero
$D_y$,
i.e. of a nonzero $s$ is roughly $\Sigma LMs^2/2$, at least for
$s\ll1$. The equipartition theorem then gives that the average
value
of $s^2$ should be of order $(NM)^{-1}$. This implies that
$E_y=0$ provided that in thermodynamic limit $N^2$ increases
faster
than the number of ledges $M$. This condition is satisfied
whenever
the ground state is described by a finite density function, with
$M/N$
finite.

For the special case at hand we have explicitly verified that
$E_y=0$: Inserting (\ref{ff:cyy}) together with
\begin{equation}
C_{yy}(0,0)=\langle 0| (w-w^{-1})\left\{ {
wc_{1}^{+}c_{1}c_{2}c_{2}^{+}+wc_{1}c_{1}^{+}c_{2}^{+}c_{2}+c_{1}
c_{2}^{+}-c_{1}^{+}c_{2}
}\right\} |0\rangle . \label{ff:c0}
\end{equation}
into \r{ey} we obtain
\bea
E_{y}(k_{F})&=&-\frac{\left\langle 0\right|
c_{2}^{+}c_{1} + c_{1}^{+}c_{2}
\left|0\right\rangle
}{w-w^{-1}}\nn
&&+\frac{1}{N^{2}}\sum_{{\rm hole}
(k_{h},+),{\rm particle}(k_{p},\alpha )}a_{+,\alpha
}(k_{h},k_{p})a_{\alpha
,+}(k_{p},k_{h})\frac{\Lambda _{+}({k_{h}})+\Lambda _{\alpha
}{(k_{p})}}{
\Lambda _{+}({k_{h}})-\Lambda _{\alpha }{(k_{p})}}\ ,
\label{ff:ey}
\eea
where we have explicitly displayed the dependence on the Fermi
momentum $k_F$. By inspection we see that
\be
E_{y}(0)=0\ . \label{ff:ey0}
\ee
Rather than trying to evaluate (\ref{ff:ey})
directly, we calculate the change
$N\delta E_{y}(k_{F})=N(E_{y}(k_{F}+\delta k_{F})-E_{y}(k_{F}))$
under a change of Fermi momentum $\delta k_{F}=2\pi N^{-1}$
\begin{eqnarray}
N\delta E_{y}(k_{F})
&=&\frac{2}{w^{-1}-w}\frac{1+e^{ik_{F}}}{s_{+,F}-s_{-,F}
} \nonumber \\
&&+\frac{1}{N}\sum\limits_{-\pi <k<\pi }\left(
a_{+,+}(k_{F},k)a_{+,+}(k,k_{F})+a_{+,+}(-k_{F},k)a_{+,+}(k,-k_{F
})\right)
\frac{\Lambda _{+}(k_{F})+\Lambda _{+}(k)}{\Lambda
_{+}(k_{F})-\Lambda
_{+}(k)} \nonumber \\
&&+\frac{1}{N}\sum\limits_{-\pi <k<\pi }\left(
a_{+,-}(k_{F},k)a_{-+}(k,k_{F})+a_{+,-}(-k_{F},k)a_{-+}(k,-k_{F})
\right)
\frac{\Lambda _{+}(k_{F})+\Lambda _{-}(k)}{\Lambda
_{+}(k_{F})-\Lambda
_{-}(k)}\ . \label{ff:deltaEy}
\end{eqnarray}
Turning the sums into integrals and integrating numerically we
find that this is indeed zero. Alternatively, if we perform the
sums
for large finite $N$ we find that the right hand side of
(\ref{ff:deltaEy}) scales like $1/N$. In conjunction with
(\ref{ff:ey0}) this shows that $E_{y}$ is indeed zero.

\section{The case of negative kink energy}

In this appendix we discuss the case $w>1$.

This corresponds to the
following extensions of the regions discussed earlier:
Region~(1a) with
$-2\pi <\omega <-2\gamma $;
Region~(1b) with $-2\pi <\omega <-2\pi
+2\gamma $; Region~(2)
with $\omega <-2\gamma $;
Region~(3) with $\omega >2\gamma $. In all
cases the
one-particle spectrum is quite different from the results
obtained for $w<1$.
In addition to the lines in spectral parameter space discussed
above, we
find that there are curves on which the bare momentum is real and
the bare
energy has a constant real part but acquires a variable imaginary
part (whereas both
bare momentum and bare energy remain real on the straight lines).

\begin{figure}[ht]
\begin{center}
\noindent
\epsfxsize=0.8\textwidth
\epsfbox{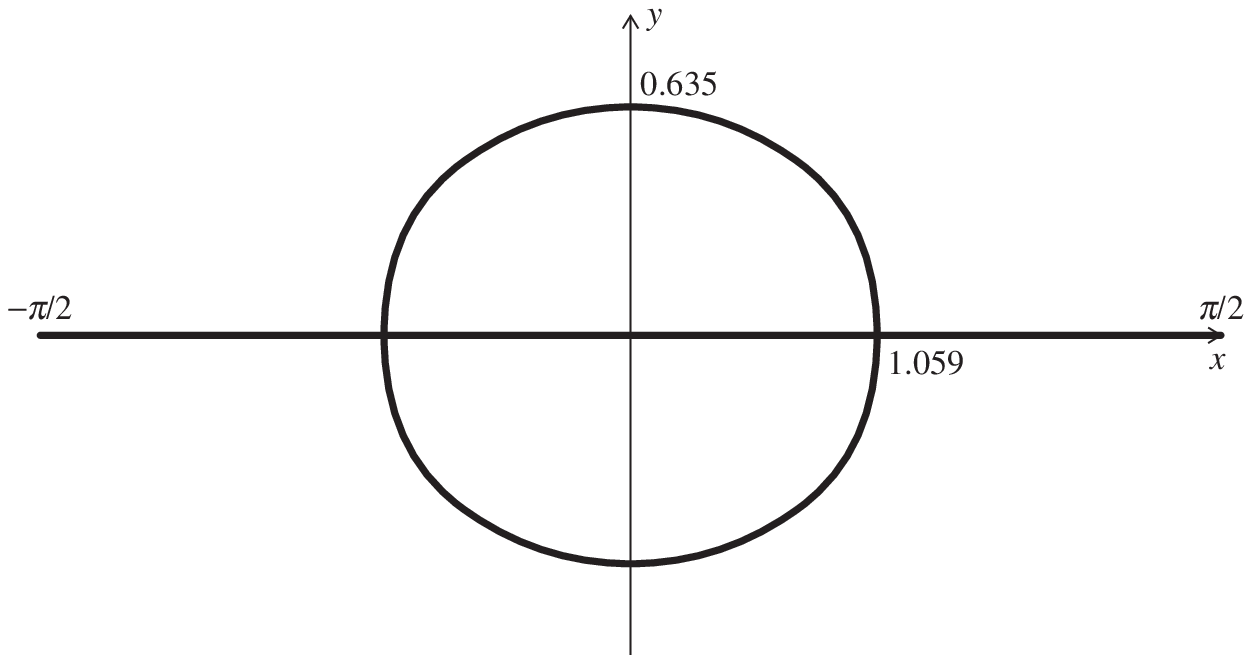}
\end{center}
\caption{\label{fig:arcs}
Location of one-strings in the $(x,y)$ plane with
$\lambda =x+iy+i\omega /4$, for $\gamma=1$ and $\omega=5$.}
\end{figure}

We specialize the remainder of this discussion to Region~(3).
The curves, shown on Fig.~\ref{fig:arcs},
are parametrized by $\lambda =x+iy+i\omega /4$ with
$x$ and $y$ real and satisfying:
\be
\cosh \gamma \cosh (2y)=\cosh(\omega /2)\cos (2x).
\label{curve}
\ee

The line corresponds to low-momentum states
$-p_c<p_0<p_c$, and the 
curves to high-momentum states $|p_0|>p_c$
($p_c$ is comprised between $0$ and $\pi$ and depends
on the parameters $\gamma$ and $\omega$).
If we consider the
dispersion relation, i.e. the (bare) energy
as a function of (bare) momentum, there
are branch points which correspond to the contact
point between the line and the curves in term
of the spectral parameter. For the high-momentum
states the bare energy becomes complex, and its
real part is constant on the whole, equal
to the energy at the contact point.

We first describe the maximum-eigenvalue state.
We believe that this state is made up of particles
with spectral parameters $\lambda_k=x_k+i\omega/4$,
where all $x_k$ are real. However the situation
is different from that considered in the main
text in one important aspect: for $\omega>2\gamma$
we find that the bare momentum is not a
monotonous function of $x$, so that the
bare density $\rho_0(x)=\frac{1}{2\pi}
dp_0/dx$ is not positive: it turns out
that its average value vanishes.

The logarithm of the Bethe Ansatz equations
is:
\be
Np_0(x_k)-\sum_j\theta(x_k-x_j)=\pi+2\pi I_k
\ee
where
\be
e^{i\theta(x)}=\frac{\sinh(x+i\gamma)}{\sinh(x-i\gamma)}
\ee
and the $I_k$ are integers, which
depend on the cut structure used for $\theta$.
We make the following choice: $\theta(x)$ is continuous
throughout the interval $-\pi<x<\pi$, with $\theta(-\pi)=2\pi$,
$\theta(0)=0$, $\theta(\pi)=-2\pi$.

A direct numerical solution of these equations for small systems
suggests that for this choice of cut structure, and
provided the spectral parameters are ordered ($x_k<x_k+1$),
then for the state corresponding to
the largest eigenvalue of the transfer matrix
we have $I_{k+1}-I_k=1$.

If we assume
that this structure is correct, and take the thermodynamic
limit, we find that the density of particles
$\rho(x_k)=(N(x_{k+1}-x_k))^{-1}$ satisfies the
following integral equation:
\be
\rho (x)+\frac{1}{2\pi }\int\nolimits_{-x_F}^{x_F}
K(x-y)\rho(y)dy=\rho _{0}(x),
\ee
where the kernel is given by:
\be
K(x)=\frac{d\theta }{dx}=\frac{-2\sinh 2\gamma }{\cosh 2\gamma
-\cos 2x}.
\ee

Since the bare energy is negative for any
value of $x$ we assume that the Fermi
level is $x_F=\pi/2$ for $v=0$.
We can therefore solve the integral equation
by using Fourier decompositions.
The result for the density is then:
\be
\rho (x)=\frac{1}{\pi }\frac{\sinh (\omega /2-\gamma )}
{\cosh(\omega /2-\gamma )-\cos(2x)}
\ee

We have performed further
numerical calculations
in order to confirm this result.

First, for systems up to size 16
(i.e. $N=8$) we have calculated directly the
largest eigenvalue of the transfer matrix
using an iteration method. This gives a good
approximation of the ground state energy per site,
and the result agrees with the analytical expression
obtained above.

For much larger systems (several hundred sites) we have
solved the Bethe Ansatz equations numerically, using
the distribution of spectral parameters predicted
by our theoretical result for the density as ``seed''
for a Broyden algorithm. The agreement obtained
by this technique confirms that the integral equation
we solve does indeed correspond to the thermodynamic
limit of the Bethe Ansatz equations.

Let us comment briefly on ``excited states'', i.e. eigenvectors of the
transfer matrix corresponding to subleading eigenvalues. One type of
excitation corresponds to removing a particle from the Fermi sea
without introducing a momentum. Its energy vanishes for $v=-\gamma/2$.

Other excitations can be constructed by creating particle-hole
pairs. However, any eigenstate of the transfer matrix must have a
real momentum. This forces us to introduce two particles at spectral
parameters $\lambda_p=x_p \pm iy_p+i\omega/4$ with $(x_p,\pm y_p)$ on
the curve \r{curve}. The two holes have spectral parameters
$\lambda_i=x_i+i\omega/4$ with $x_i$ real for $i=1,2$. The momentum
and energy of this excited state is found to be
\bea
e^{ip}&=&e^{-4i(2x_p-x_1-x_2)} \frac{f(x_p+iy_p)f(x_p-iy_p)}
{f(x_1)f(x_2)}\ ,\\
e^{-\delta E}&=&
\frac{
|\sin(x_1+i\omega /4-i\gamma/2)\sin(x_2+i\omega /4-i\gamma/2)|^2
}{|\sin(x_p+iy_p+i\omega /4-i\gamma/2)\sin(x_p+iy_p-i\omega /4+i\gamma/2)|^2
},
\eea
where
\be
f(x)=\frac{\sin(x+i\omega/4-i\gamma/2)}{\sin(x-i\omega/4+i\gamma/2)}.
\ee

\section{Second-quantized form of transfer operator}

The transfer operator which we use is a product
of noncommuting terms,
which operate on alternating sublattices.
Consider a vertex as shown in Fig.~\ref{fig:figver}
which separates two vertically-consecutive
horizontal rows of edges.  When an edge is occupied
by a ledge, let it be occupied by a particle,
or an up spin.  Then the transfer operator
for site $j$ is:
\be
T_j=\left(
1+e^{-\tau}(\sigma_j^+\sigma_{j-1}^-+
\sigma_{j+1}^+\sigma_j^-)
\right)
\left(
1+(e^b-1)n_jn_{j+1}
\right)
\ee
with $n_j=2\sigma_j^z-1$.  Since $n_j^2=n_j$, basis
states with $\sigma_j^z\sigma_{j+1}^z=1$ are annihilated
by $(\sigma_j^+\sigma_{j+1}^-+\sigma_{j+1}^+\sigma_j^-)$,
but on those with $\sigma_j^z\sigma_{j+1}^z=-1$
we find that
$(\sigma_j^+\sigma_{j+1}^-+\sigma_{j+1}^+\sigma_j^-)^2$
acts as the identity operator.  It follows that
\be
T_j=(\sinh 2\tau)^{1/2}\exp
\left(\tau^*\left(
\sigma_j^+\sigma_{j+1}^-+\sigma_{j+1}^+\sigma_j^-
\right)\right)
\exp\left(bn_jn_{j+1}\right)
\ee
which may be written in terms of Fermi operators
in the usual way.  This allows us to make contact
with the appendix of \cite{jaya}.  Note however that,
contrary to the statement implied there, $T_j$ and
$T_{j+1}$ do not commute, because the ``hopping''
terms do not allow it.
But, if we take $b$ and
$\tau^*$ so small that only linear terms
can be retained, where $\tau^*$ is defined by
$w=\tanh\tau^*$, then the transfer matrix
\be
T=\prod_1^N T_{2j}\prod_1^N T_{2j-1}
\ee
can be approximated by $T\approx (2\sinh 2\tau)^N
e^H$ with
\be
H=b\sum_1^{2N}n_j n_{j+1} + \tau^*\sum_1^{2N}
(\sigma_j^+\sigma_{j+1}^-+\sigma_j^-\sigma_{j+1}^+).
\label{ham}
\ee
This is the XXZ model in a field.  We emphasize
that \r{ham} is an uncontrolled approximation which
is not needed in our solution.

\section{Ledge-ledge interactions}

The interaction between two ledges may be investigated by
considering the incremental free energy per unit
cylinder length with $2p$ ledges (the $2p$-particle
sector, that is) on a cylinder having finite circumference
$2N$, length $L$ and joined along the cylinder
axis to form a torus.
We define the incremental free energy by
\be
f^\times(2p,N)=\lim_{L\to\infty}\frac{1}{L}
\log\frac{Z_{2p}(L,2N)}{Z_0(L,2N)}
\ee
where $Z_{2p}(L,2N)$ is the partition function
on the torus for $2p$ particles.
This is evaluated by giving the momenta
their lowest values $({\rm mod}\ 2\pi)$
consistent with \r{bae}.  Asymptotically for
large N, in that part of the parameter space
with no ledge binding, this equation
becomes
\be
e^{ik_jN}=-1
\ee
for $j=1,\dots 2p$ with minimal solutions
\be
k_j=(2j-2p-1)\pi/N.
\ee
From \r{eps0}, we have
\be
f^\times(2p,N)\approx 2p\log w +
\left(\frac{d^2\epsilon_0}{dp_0^2}\right)_{p_0=0}
\sum_1^p k_j^2.
\label{f2pN}
\ee
We assume that the second term above is made up
of $2p$ equal ledge-ledge pair interactions and
that the ledges are on average equally spaced
at a distance $l=N/p$.  Then this asymptotic pair
interaction $u(p,l)$ is given by:
\bea
u(p,l)&=&\frac{\pi^2}{2p^3l^2}
\left(\frac{d^2\epsilon_0}{dp_0^2}\right)_{p_0=0}
\sum_1^p(2j-1)^2 \nn
&=&\frac{\pi^2w}{12l^2}\frac{2p^2-1}{p^2}.
\label{upl}
\eea
Notice firstly that $u(p,l)$ is independent
of $b$.  It depends on the sector $p$; this
is an example of the effect mentioned by
Fisher \cite{fisher} in the wetting
context.  Much interest attaches to
the case with a small density $D$ of ledges,
which is the same as that given by \r{upl}
in the $p\to\infty$ limit.  In this regime the
following approximation can be used:
\be
D=2x_F\rho(0)+\rho^{(2)}(0)\frac{x_F^3}{3}+O(x_F^5)
\label{appE:D}
\ee
where
\be
\rho(0)=\left(1-K(0)\frac{x_F}{\pi}\right)^{(-1)}\rho_0(0)
+O(x_F^3)
\ee
and
\be
\rho^{(2)}(0)=\rho_0^{(2)}(0)
+\left(\frac{x_F}{\pi}\right)^2K^{(2)}(0)\rho_0(0)+O(x_F^4)
\ee
It is natural to define $y_F$ by
\be
y_F=x_F\left(1-K(0)\frac{x_F}{\pi}\right)^{-1}
\label{appE:dressed}
\ee
so that $y_F\rho_0(0)=x_F\rho(0)$, a dressing
of the Fermi level.  Thus
\be
D=2\rho_0(0)y_F+\frac{\rho^{(2)}(0)}{3}y_F^3+O(y_F^4)
\ee
which is readily inverted, giving
\be
y_F(D)=\frac{1}{2\rho_0(0)}D-\frac{\rho^{(2)}(0)}{48\rho_0(0)^4}
D^3+O(D^4)
\ee
The free energy is obtained from \r{Egs2} by approximating
the small integral for small $x_F$ and then inserting
\r{appE:dressed}, leading to
\bea
E&=&2a
+2y_F\rho_0(0)\epsilon_0(0)
+\frac{y_F^3}{3}\left(
\epsilon_0^{(2)}\rho_0(0)+\epsilon_0\rho_0^{(2)}(0)
\right)+O(y_F^4)
\nonumber \\
&=&2a+\epsilon_0(0)D
+\frac{D^3}{24}\frac{\epsilon_0^{(2)}(0)}{\rho_0(0)^2}
+O(D^4)
\eea
Now
\be
\frac{\epsilon_0^{(2)}(0)}{\rho_0(0)^2}=2\pi^2 w
\ee
giving
\be
E=2a+\epsilon_0(0)D+\frac{D^3\pi^2w}{12}+O(D^4).
\ee

A similar expansion can be used near the
transition between the massless
and ferromagnetic phases.  This transition
can be driven either by the chemical potential
or by temperature.  In both cases the analytic
behaviour near the transition can be studied using
the parameter $t=\epsilon_0(0)$, which is equal
to $a-a_c$ if the relevant parameter is the chemical
potential $a$, and $t\propto T-T_c$ if using the
temperature.  The Fermi level is obtained by
expanding the left-hand side of $\epsilon(x_F)=0$:
to leading order $\epsilon(0)=\epsilon_0(0)$ and
$\epsilon^{(2)}(0)=\epsilon^{(2)}_0(0)$, therefore
$x_F\propto\sqrt{t}$.  It follows that the free
energy has a $t^{3/2}$ singularity, and thus that the
specific heat diverges like $t^{-1/2}$.

\end{document}